\documentclass[floatfix,amsmath,amssymb,superscriptaddress,longbibliography,onecolumn,showpacs,nobibnotes]{article}
\usepackage{amssymb}
\usepackage{geometry}
\usepackage{amsmath}
\usepackage{bbold}
\usepackage{bm}

\usepackage{braket}
\usepackage[title]{appendix}
\usepackage{xcolor}
\usepackage{soul}
\usepackage{graphicx}
\usepackage{textcomp}
\usepackage{enumerate}
\usepackage[colorlinks=true, linkcolor=red, citecolor=cyan, urlcolor=blue]{hyperref}
\usepackage{cleveref}
\usepackage{lipsum, babel}
\graphicspath{{../figura1/}{../figura2/}}
\usepackage{graphicx}
\newcommand*\Tr{\mathop{}\!\mathrm{Tr}}

\usepackage{soul}

\usepackage{xcite}
\usepackage{xr}
\usepackage{authblk}
\makeatletter

\date{}

\title{Statistical mechanics of transfer learning in fully-connected networks in the proportional limit}
\author{Alessandro Ingrosso \thanks{ingrosso@ictp.it}}
\affil{The Abdus Salam International Centre for Theoretical Physics (ICTP), Trieste, Italy}

\author{Rosalba Pacelli}
\affil{I.N.F.N., sezione di Padova, Via Marzolo 8, 35131, Padova, Italy}

\author{Pietro Rotondo}
\affil{Dipartimento di Scienze Matematiche, Fisiche e Informatiche,
Università degli Studi di Parma, Parco Area delle Scienze, 7/A 43124 Parma, Italy}

\author{Federica Gerace \thanks{federica.gerace@unibo.it}}

\affil{Dipartimento di Matematica, Università di Bologna,
Piazza di Porta San Donato 5, 40126, Bologna (BO), Italy}

\begin{document}
\maketitle
\begin{abstract}
    Transfer learning (TL) is a well-established machine learning technique to boost the generalization performance on a specific (target) task using information gained from a related (source) task, and it crucially depends on the ability of a network to learn useful features. Leveraging recent analytical progress in the proportional regime of deep learning theory (i.e. the limit where the size of the training set $P$ and the size of the hidden layers $N$ are taken to infinity keeping their ratio~$\alpha = P/N$ finite), in this work we develop a novel single-instance Franz-Parisi formalism that yields an effective theory for TL in fully-connected neural networks. Unlike the (lazy-training) infinite-width limit, where TL is ineffective, we demonstrate that in the proportional limit TL occurs due to a renormalized source-target kernel that quantifies their relatedness and determines whether TL is beneficial for generalization.
\end{abstract}

\section{Introduction} 
Modern deep learning relies on foundation models that are pre-trained on tasks closely related to the one of interest but much richer in training examples. In this way, the generalization performance of a neural network trained on a data-scarce \emph{target task} can consistently improve by leveraging the knowledge that the pre-trained model has previously acquired on a close but data-abundant \emph{source task}. This \emph{Transfer Learning} (TL) practice has amply demonstrated to enhance the generalization performance of deep learning models, especially in those settings where data is scarce or labelling is demanding~\cite{bernhardt2022active,mincu2022developing}.

Despite being among the dominating paradigms in deep learning applications, TL remains poorly understood from a theoretical perspective, with several fundamental questions still open. For instance, (i) \emph{how does the source-target similarity affect TL efficiency?} (ii) \emph{how does the width of the transferred layers impact generalization performance?}

Most theoretical results in this direction hold for a parallel form of TL in the framework of classical learning theory, and rely on proofs of worst-case bounds, based on the Vapnik-Chervonenkis dimension~\cite{vapnik2013nature}, covering number, stability, and Rademacher complexity~\cite{bartlett2002rademacher} (see~\cite{zhang2022transfer,wang2020generalizing,zhang2021survey} for review). A recent line of research has approached TL using statistical mechanics~\cite{lampinen2018analytic,dar2022double,dhifallah2021phase}. However its applicability is limited, since its focus is on linear NNs and source-target data models are overly simplistic.
In~\cite{gerace2022probing}, the authors went one step further by proposing a theoretical framework to study TL in one-hidden layer (1HL) and non-linear networks. Here, pre-training is purely numerical, the interaction between the source and target is encoded implicitly in the empirical covariance of the hidden units, and the first layer weights are always kept fixed to the source configurations.

The analytically tractable lazy-training infinite-width limit~\cite{Neal, LeeGaussian, JacotNTK,  g.2018gaussian, Hanin2023}, one of the recent milestones in deep learning theory, is also not a viable option to theoretically investigate pre-training and transfer stages: since the statistics of the weights remains unchanged during training, no features can be transferred from the source to the target task~\cite{tensoriv}. One possibility to overcome this issue is to consider the feature-learning phase of infinite-width networks~\cite{doi:10.1073/pnas.1806579115,  NEURIPS2022_d027a5c9, https://doi.org/10.1002/cpa.22074, doi:10.1137/18M1192184,NEURIPS2018_a1afc58c, seroussi2023natcomm, NEURIPS2021_b24d2101}, where TL is still possible~\cite{yang2020feature}. 
To investigate more realistic settings than the infinite-width limit, one could analyze TL in the recently-explored proportional regime of deep NNs, formally defined as the limit where both the size of the training set $P$ and the width of the hidden layers $N_\ell$ ($\ell = 1, \dots, L$, $L$ being the depth of the network) 
scale to infinity while keeping the ratios $\alpha_\ell = P/N_\ell$ fixed. This limit has been firstly studied in linear networks~\cite{SompolinskyLinear, doi:10.1073/pnas.2301345120, bassetti2024feature,tiberi2024dissecting} and then extended to non-linear models~\cite{pacelli2023statistical, aiudi2023, li2022globally}. One major advantage of such a setting is the possibility of corroborating its analytical predictions with the outcome of Bayesian learning experiments in finite-width networks, as recently done for fully-connected 1HL models~\cite{baglioni2024predictive}. 
\begin{figure}
    \centering
    \includegraphics[width=1\textwidth]{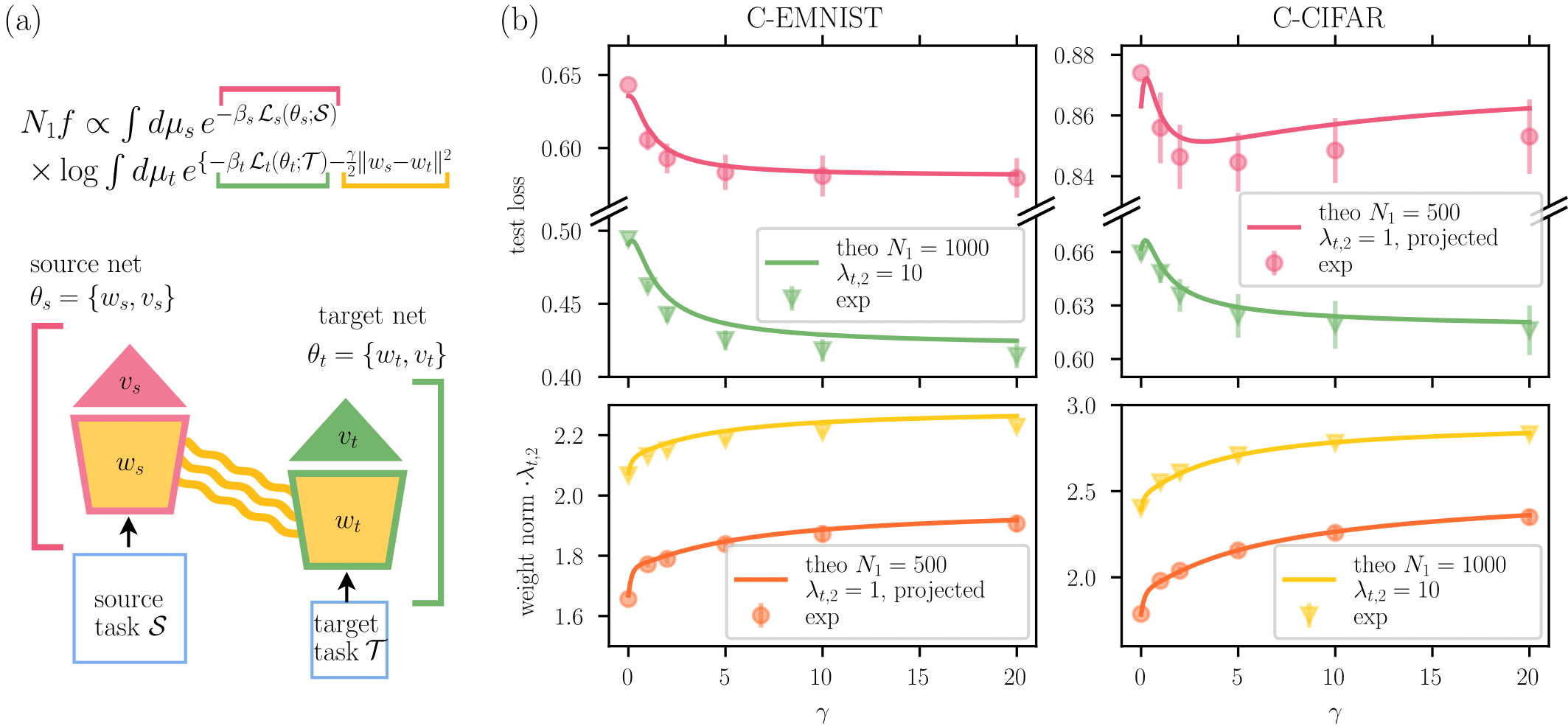}
    \caption{\textbf{Transfer learning in one-hidden layer networks through the lens of a single-instance Franz-Parisi framework.}  \textbf{(a) Sketch of transfer learning with 1HL networks.} The first-layers weights of the source (left) and target (right) networks are coupled, while the readout layers are optimized independently. The coupling strength is modulated by the parameter $\gamma$ (wavy yellow lines). The vertical displacement between the source and target networks represents the temporal displacement in their training, physically implemented by quenched averaging of the free-energy (equation above).
    \textbf{(b) Learning curves of target 1HL networks.} Experimental test loss (top) and squared norm of last layer weights (bottom) are shown as a function of source/target interaction strength $\gamma$ for different sizes $N_1$ and magnitudes of Gaussian prior $\lambda_{t,2}$ of the last layer. Markers are shown in comparison to solid lines, which represent theoretical predictions from the theory. The networks are trained on $P_{s} = 800$, $P_{t} = 100$ examples from 
    C-EMNIST (left) and C-CIFAR (right) tasks
    (see Appendix for more details on the tasks). Bars represent one standard deviation. The curves at $N_1 = 1000$ were obtained with source networks pre-trained on  $784$-dimensional examples. The curves with $N_1 =500$ refer to data projected in $D=300$ dimensions.
      }
    \label{fig:TL_with_SIFP}
\end{figure}

In this manuscript, we leverage the results established in the proportional limit, combining this approach with a novel \emph{single-instance Franz-Parisi} formalism~\cite{franz1995recipes} to investigate TL effectiveness in Bayesian NNs. In particular, we argue that the posterior over the weights of the source task plays the role of the quenched disorder in spin-glass theory~\cite{mezard1987spin}, which can be thus integrated out using the well-known replica method. This leads to an explicit formula for the free-energy in the proportional limit, describing the learning scenario of a neural network trained on the target task while coupled to a quenched copy, pre-trained on the source task.

\section{Single-instance Franz-Parisi formalism for Transfer Learning}
In the standard TL pipeline, a neural network is trained on a $P_t$-dimensional target set $\mathcal T = (X_t, y_t)$ while keeping some of its layers frozen to the ones transferred from the $P_s$-dimensional source set $\mathcal S = (X_s, y_s)$. Deep learning practitioners can later add a fine-tuning stage, where the transferred layers are unfrozen and the whole network is trained on the target set using a smaller learning rate.

To rationalize the effectiveness of TL in the proportional limit of fully-connected networks, we introduce a novel approach based on statistical mechanics of learning \cite{engel2001statistical}. Specifically, we consider a setting involving a one-hidden layer neural network $\phi_t$ whose first-layer weights adapt to the target task while coupled to those learned by another network $\phi_s$ on the source task. A sketch of the learning setting is shown in Fig.~\ref{fig:TL_with_SIFP}a. In the framework of statistical mechanics, this learning paradigm can be effectively described by the following free-energy density:
\begin{equation}
f = \frac{1}{N_1}\mathbb{E}_{\theta_s} \left[
  \log\int d \mu(\theta_{t})e^{-\beta_{t}\mathcal{L}_{t}\left(\theta_{t}; \mathcal T \right)-\frac{\gamma}{2} \left\Vert w_{s}-w_{t}\right\Vert ^{2}}\right],
 \label{eq:F_1HL}
\end{equation}
where $N_1$ is the number of hidden units, $\theta_{s/t}$ are the collections of the first and second-layer weights $w_{s/t}$ and $v_{s/t}$ of $\phi_{s/t}$ respectively, $\mu (\theta_{t})$ is the Gaussian prior over the target weights, $\mathcal L_t$ is the target training loss, and $\beta_{t}$ is the target inverse temperature. The limit $\beta_{t} \to \infty$ can eventually enforce perfect interpolation of the target training set. The coupling between the source and the target network is controlled by a parameter $\gamma$. This allows for continuous interpolation between two regimes: one where the network is trained from scratch on the target set with no knowledge transfer from the source task ($\gamma = 0$), and another one where the first-layer weights of the target network are kept frozen to the source weights, while the second-layer weights adapt to the target set ($\gamma \to \infty$).
The intermediate values of $\gamma$ describe the fine-tuning stage, where the first-layer weights $w_t$ are first initialized to the source configuration $w_s$ and then trained on the target set, together with the second-layer ones $v_t$. The expectation over the source configurations $\theta_s$ ensures we describe the typical TL behavior.

To guarantee that the source configurations effectively solve the source task, we take this expectation over the \emph{posterior} distribution of the source weights. This corresponds to a Boltzmann-Gibbs measure whose partition function only involves the source task:
\begin{equation}
    Z_{s}\left(\beta_{s}\right)=\int d\mu\left(\theta_{s}\right)e^{-\beta_{s}\mathcal{L}_{s}\left(\theta_{s}, \mathcal S\right)},
\end{equation}
where $\mu (\theta_{s})$ is the Gaussian prior over the source weights and $\beta_{s}$ is the source inverse temperature. It is important to emphasize that, since the source and target training are not performed simultaneously, the expectation over the source configurations is quenched. This is crucial to ensure that the source posterior is not affected by the source-target coupling: in a TL pipeline, the source network is indeed trained on the source task without relying on any information on the target data.
 
As anticipated, the free-energy density in Eq.~\eqref{eq:F_1HL} closely resembles the Franz-Parisi potential, originally introduced to analyze metastable states in spin-glass systems~\cite{franz1995recipes} and later used to characterize the properties of energy landscapes in machine learning problems~\cite{baldassi2015subdominant,baldassi2016learning} or knowledge distillation~\cite{saglietti2022solvable} and curriculum learning~\cite{saglietti2022analytical}. 

At variance with previous literature~\cite{pmlr-v202-cui23b,camilli2023fundamental}, we refrain from averaging the free-energy density over the input data distribution, in the same spirit of Refs. ~\cite{SompolinskyLinear, pacelli2023statistical, li2022globally, aiudi2023, seroussi2023natcomm} (note that a similar approach has been put forward to investigate Markov proximal learning~\cite{avidan2023connecting}, in an attempt to connect the neural tangent and the neural network Gaussian process kernels). 

For this reason, we name our new theoretical framework \emph{single instance} Franz-Parisi. The quenched expectation over the source weights is tackled using the replica method, while the integral over the replicated target configurations in Eq.~\eqref{eq:F_1HL} is performed via the standard kernel renormalization approach, which is exact for deep linear networks~\cite{SompolinskyLinear, li2022globally}, and can be justified using a Gaussian equivalence for non-linear activation function. In the following, we provide a sketch of the derivation [full details can be found in the Appendix], which is valid for quadratic loss function (mean squared error). 

\subsection{Free-energy in the proportional limit} To evaluate the quenched free-energy in Eq.~\eqref{eq:F_1HL} in the proportional thermodynamic limit described in the introduction, we make use of the replica trick:
\begin{equation}
f =\frac{1}{N_{1}}\frac{1}{Z_{s}\left(\beta_{s}\right)}\lim_{n\to0}\partial_{n}Z^{n}.
\end{equation}
The calculation yields a compact expression for the replicated partition function in terms of an effective finite-$n$ action $S_n$, which, for the sake of simplicity and lighter notation, we here show in the case $\beta_t=\beta_s=\beta$ and $\lambda_{s,2} = \lambda_{t,2} = \lambda$:
\begin{align}
    Z^{n}&\sim \mathrm{exp} \left\{N_{1} S_n(\mathcal Q, \bar{\mathcal{Q}};n)\right\}  \\
S_n &= \mathrm{Tr} \left(\lambda \mathcal{Q}\mathcal{\bar{Q}}\right)-\log\det\left(\mathbb 1+\mathcal{\bar{Q}}\right)-\frac{1}{N_{1}}\log\det \left(\mathbb{1}+ \beta \mathcal{K}\right)-\frac{\beta}{N_{1}}y^{T}\left(\mathbb{1}+\beta\mathcal{K}\right)^{-1}y
\end{align}
where $\mathcal{Q}$ and its conjugate $\bar{\mathcal Q}$ are $(n+1) \times (n+1)$ order parameters matrices (see Appendix for the most general expression).
At this level, $\mathcal{K}$ is a $\left(P_s+nP_t\right) \times \left( P_s+nP_t\right)$ replicated renormalized kernel matrix:
\begin{align}
    & \mathcal{K}_{\mu \nu}^{ab} =  \mathcal{Q}^{ab} K^{ab}_{\mu \nu},  \qquad \,\, K^{ab}_{\mu\nu}=\left\langle \sigma\left(h_{\mu}^{a}\right)\sigma\left(h_{\nu}^{b}\right)\right\rangle ,
\end{align}
where the expectations are taken over an effective distribution of the first-layer hidden representations $h$, and where the $0$-th replica identifies the source network. Given the replica-symmetric nature of this problem, there are only four distinct order parameters $ \mathcal Q = \{ Q_s, Q_t, Q_{st}, Q_{tt}\}$ -- along with their conjugates $\bar{\mathcal{Q}}$ -- mediating the interaction between the source and the target posterior distributions, via their coupling with four distinct kernels $K_s$, $K_t$, $K_{st}$, $K_{tt}$. Of these, the crucial one is the effective source-target kernel $K_{st}$, computed on Gaussian pre-activations with covariance that depends on the overlap matrix between the inputs in the source and target task $C_{st}={X_{s}X_{t}^{T}}/{N_0}$, with $N_0$ being the input dimension.

The $\mathcal O (1)$ terms of $S_n$ conspire to rebuild the effective action of the source task alone~\cite{pacelli2023statistical}, while the $\mathcal O(n)$ terms in $S_n$ in the $n \to 0$ yield the genuine source-target action, incorporating the effect of transfer. Through appropriate derivatives of such \emph{transfer action}, we can compute relevant observables in the equilibrium ensemble, e.g. the test loss or the statistics of the last-layer weights $\langle v^2 \rangle$ (see Appendix).

\begin{figure}
    \centering
    \includegraphics{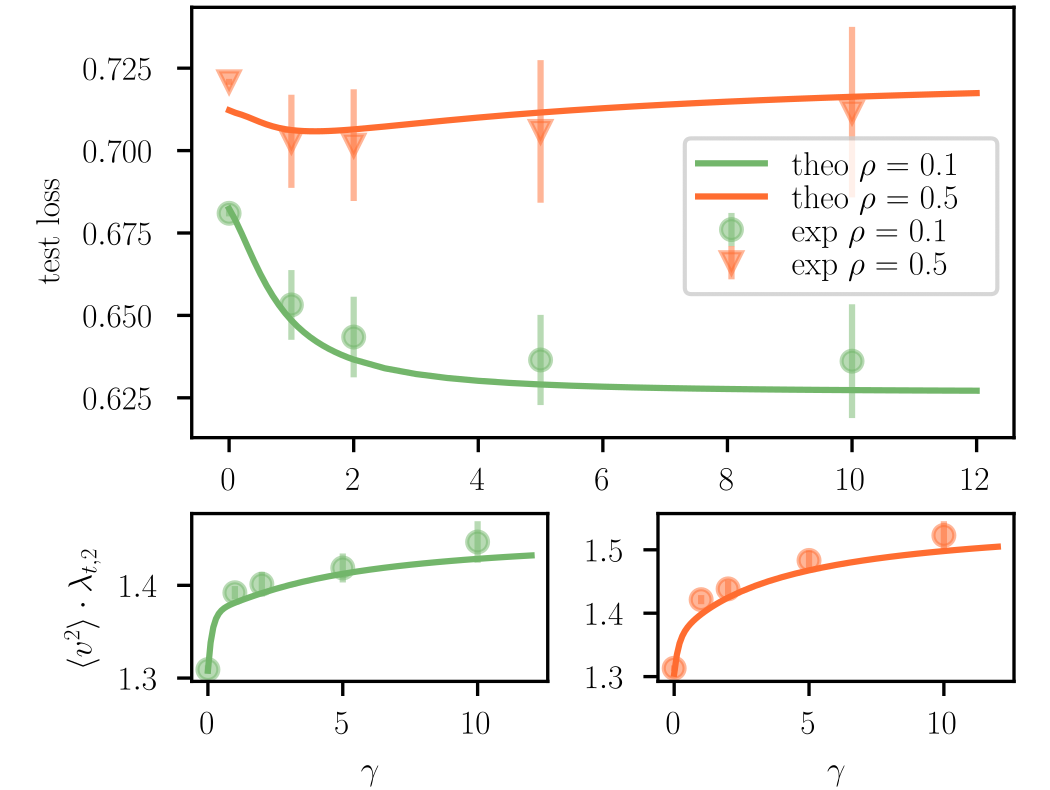}
    \caption{\textbf{Effectiveness of fine-tuning is linked to correlation between source/target tasks.} Top panel: test loss of 1HL target network is shown as a function of interaction strength with source network $\gamma$ for two values of correlation between source/target $\rho = 0.1, 0.5$. The source task comprises $P_{\textrm{source}} = 800$ examples generated with a hidden manifold model with $N=500$ and latent dimension $D=300$. The $P_{\textrm{target}} = 200$ target examples are build using a fraction $1-\rho$ of the source features, and a fraction $\rho$ of new ones.
    Middle and bottom panels: the order parameter $Q$ is shown as a function of $\gamma$ respectively for $\rho = 0.1 $ and $\rho = 0.5$. }
    \label{fig:fig2_hmm}
\end{figure}

\section{TL on benchmark tasks}
Fig.~\ref{fig:TL_with_SIFP}b illustrates the good agreement between theoretical predictions (solid lines) and numerical simulations (dots) for both the test loss (first row) and the norm of the last-layer weights (second row), as a function of the coupling parameter $\gamma$, for two different values of $N_1$ and last-layer regularization strength $\lambda_{t,2}$. The two columns refer to TL scenarios with two different source-target pairs of binary tasks, namely C-EMNIST (left) and C-CIFAR (right), built from the well-known benchmark computer vision datasets EMNIST and CIFAR10. In particular, in the same spirit of~\cite{gerace2022probing}, we build the source task by first dividing some classes of the original dataset into two groups, and then assigning one label per group. The target task is then obtained from the source task by changing one class per group. In this way, some of the features characterizing the source data are absent in the target set. This case is thus meant to describe settings where the source and target tasks are related but differ because of style or geometric structure of the input data~\cite{gerace2022probing}.

The effectiveness of TL and fine-tuning strongly depends on data structure, and thus on the specific source-target relation, but also on the network size, as shown in Fig~\ref{fig:TL_with_SIFP}b. For instance, while there exists an optimal $\gamma$ minimizing the test loss in the C-CIFAR setup (right pink curve), the same does not occur with C-EMNIST (left pink curve), where the generalization performance steadily improves with $\gamma$, eventually reaching a plateau for $\gamma \to \infty$. At larger network sizes, the optimum in the C-CIFAR setup (left green curve) is attained at very large $\gamma$: fine-tuning is thus less beneficial than just freezing the features when converging towards the infinite-width limit. In the next two sections, we will focus on the impact of data structure and network width on TL.

\subsection{Fine-Tuning and source-target similarity}
The level of source-target similarity is a crucial aspect in TL settings. If two tasks are poorly related, it is not unusual to observe negative transfer effects~\cite{gerace2023optimal}, up to the point where fixing the model parameters at random is more convenient than transferring those learned on the source task~\cite{gerace2022probing}. To analyze these aspects, we use the \emph{correlated hidden manifold} (CHMM), a synthetic data model where source-target correlations are tuned via a set of parameters meant to mimic different and realistic TL scenarios. For instance, the source-target datasets may differ because of structure and style, as in the example in Fig.~\ref{fig:TL_with_SIFP}. In the model, this is described by the parameter $\rho$, which controls the fraction of source features that are replaced by new ones in the target set (more details on the CHMM can be found in Appendix).

Fig.~\ref{fig:fig2_hmm} shows the test loss (top row) and norm of the weights (bottom row) as a function of $\gamma$, when training on a CHMM source-target pair with two distinct values of $\rho$.  
When a larger number of features are common to both tasks (green curve), the test loss decreases with the strength of the source-target coupling, showing that it is always convenient to constrain the first-layer weights to the source ones ($\gamma \to \infty$) rather than training from scratch ($\gamma$ = 0) or slightly fine-tuning the network on the target set (small $\gamma \to 0$). Instead, when the two tasks share only half of the features (orange curve), one clearly sees that freezing the first-layer weights
does not lead to better generalization performance than training from scratch. In this case, a slight improvement is only attained by fine-tuning the network on the target set at $\gamma \simeq 1$.

\subsection{TL at proportional width VS infinite-width}
\begin{figure}
    \centering
    \includegraphics{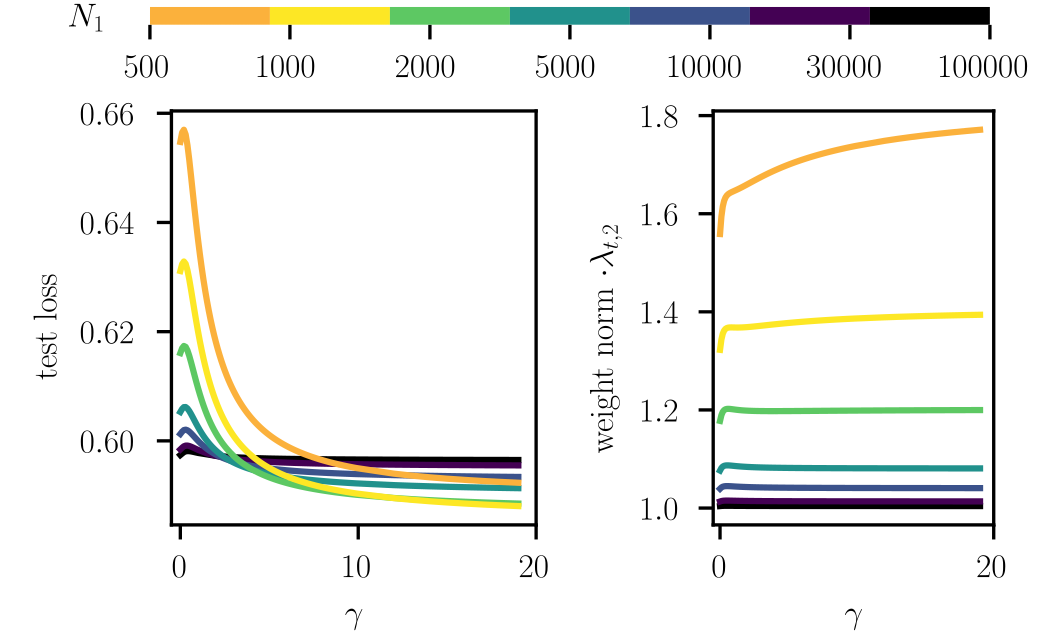}
    \caption{\textbf{Transfer learning is ineffective in the lazy-training infinite-width limit.} Predicted test loss (left) and norm of last-layer weights (right) of the target network are shown as a function of the interaction strength $\gamma$ for different values of $N_1$. The task pair is C-EMNIST with $P_s = 800$, $P_t = 100$. As the infinite-width limit is approached ($N_1 \gg P_s, P_t$ ), TL becomes ineffective, with no gain in terms of performance between the uncoupled model ($\gamma = 0$) and the TL model ($\gamma >0$). Coherently, as $N_1$ grows, the last layer average weight norm converges to its prior value $1/\lambda_{t,1}$, corresponding to the infinite-width lazy-training solution $\mathcal Q = \mathbb 1$ ($Q_s = Q_t = 1$, $Q_{st} = Q_{tt} = 0$). }
    \label{fig:fig3_infwidth}
\end{figure}

Fig.~\ref{fig:fig3_infwidth} shows a comparison between the single-instance Franz-Parisi in the proportional regime and the infinite-width one. Specifically, we show test losses (left) and last-layer weight norms (right) of target networks for different values of $N_1$. Consistently, the theory reproduces the infinite-width limit behaviour in the regime where $N_1 \gg P_s ,\,  P_t$. Already at $N_1 \sim 10 P_s$, there is less than $ 2\%$ gain in terms of generalization performance between uncoupled model to the best (completely transferred) model, and the last-layer weights have the same average norm as before the training ($1/\lambda_{t,2}$). This behavior is compatible with the lazy-training infinite-width regime, where the statistics of the weights of the source and target networks does not change during training, forbidding TL from being beneficial. More interestingly, already at $\gamma \sim 1$ smaller architectures do surpass the infinite-width performance, which is the best one for the uncoupled model. The fact that finite coupled networks outperform their best uncoupled predictor signals that the performance improvement is genuinely due to transfer and is not an artifact of effective regularization.

\section{Discussion and Conclusions}
In this work, we introduced a new theoretical approach to study TL in the proportional limit that leverages techniques used in the theory of spin glasses and kernel methods, and showcased it for 1HL fully connected networks. Our theory predicts the emergence of a source-target kernel $K_{st}$, whose renormalization in the proportional limit captures the improvement on generalization performance of the target network, when source and target tasks share some structure. More interestingly, the effect of transfer allows target networks to outperform the best predictor of the uncoupled models.

A transfer pipeline can also be formulated in convex problems where there is a strong imbalance between available data for two different learning tasks: linear regression tasks is arguably the simplest example of such a case where analytical expressions for TL can be derived without resorting to the replica method (see Appendix C and fig.~\ref{fig:Transfer-regression} for a simple example in a teacher-student setting).
The generalization to deep networks and more complex layer structures -- where kernel renormalization has been shown to depend on a local mechanism~\cite{aiudi2023} -- are interesting avenues for future work we are currently pursuing. In particular, we show in the Appendix how kernels would change for a convolutional layer, and sketch a tentative derivation for multi-layer architectures, which we conjecture to be exact in the case of deep linear networks.

Finally, it is reasonable to expect that the proposed effective theory breaks down in the large $P$ regime~\cite{li2024representations, pmlr-v202-cui23b, aguirrelopez2024randomfeaturespolynomialrules}, i.e. when the number of training patterns is proportional to the total number of parameters of the network. Addressing this regime can be considered a major challenge for scientists working in deep learning theory.

\subsubsection*{Aknlowledgements}
P.R. is supported by \#NEXTGENERATIONEU (NGEU) and funded by the Ministry of University and Research (MUR), National Recovery and Resilience Plan (NRRP), project MNESYS (PE0000006) “A Multiscale integrated approach to the study of the nervous system in health and disease” (DN. 1553 11.10.2022). R.P. and P.R. thank Paolo Baglioni for fruitful discussions.

\bibliographystyle{unsrt}
\bibliography{references}
\clearpage
\textbf{\huge{Appendix}}
\appendix
\section{Setting and notation\label{subsec:Setting-and-notation}}

The output of a one-hidden layer fully connected network given a data point $x\in R^{N_{0}}$
is:
\begin{equation}
\phi\left(x\right)=\frac{1}{\sqrt{N_{1}}}\sum_{k=1}^{N_{1}}v_{k}\sigma\left(\sum_{i=1}^{N_{0}}w_{ki}x_{i}\right)
\end{equation}
with $\sigma$ a non-linear function. We will use erf as an example
of a symmetric, saturating function: generalization to other nonlinearities
is straightforward. Given a dataset of inputs and outputs $\mathcal{D}=\left\{ X,y\right\} $,
with $X$ the usual design matrix $X=\left(X\right)_{\mu i}$, the
loss function reads:
\begin{equation}
\mathcal{L}\left(\theta,\mathcal{D}\right)=\frac{1}{2}\sum_{x,y\in\mathcal{D}}\left[y-\phi\left(\theta,x\right)\right]^{2}
\end{equation}
where $\theta\equiv\left\{ w,v\right\} $ is a shorthand for the collection
of first and second layer weights.

We consider the learning problem of a one-hidden layer \emph{target}
network whose first-layer weights are coupled via a parameter $\gamma$
to those of a previously trained \emph{source} network. Let us consider
three datasets $X_{s}$, $X_{t}$, $X_{\tau}$ of size $P_{s}$, $P_{t}$,
$P_{\tau}$, respectively the training set for the source and target
task, and test set for the target task, with their respective outputs
$y_{s}$, $y_{t}$, $y_{\tau}$.

Performing the quenched average over the source posterior weights
of the log-partition function of the transfer weight, we get to the
following expression:

\begin{equation}
N_{1}f=\frac{1}{Z_{s}}\int d\mu_{s}\left(w_{s}\right)e^{-\beta_{s}\mathcal{L}_{s}\left(w_{s}\right)}\log\int d\mu_{t}\left(w_{t}\right)e^{-\beta_{t}\mathcal{L}_{t}\left(w_{t}\right)-\frac{\gamma}{2}\left\Vert w_{s}-w_{t}\right\Vert ^{2}}
\end{equation}
where
\begin{equation}
d\mu_{s/t}\left(w\right)=dwe^{-\frac{\lambda_{s/t,1}}{2}\left\Vert w\right\Vert ^{2}-\frac{\lambda_{s/t,2}}{2}\left\Vert v\right\Vert ^{2}}
\end{equation}
and
\begin{equation}
Z_{s}\left(\beta_{s}\right)=\int d\mu_{s}\left(w_{s}\right)e^{-\beta_{s}\mathcal{L}_{s}\left(w_{s}\right)}
\end{equation}
is the source partition function. To deal with the $\log$, we make
use of the replica trick:
\begin{equation}
\log Z=\lim_{n\to0}\partial_{n}Z^{n}
\end{equation}
and we write the transfer free-entropy in terms of replicated variables as
\begin{align}
N_{1}f & =\frac{1}{Z_{s}\left(\beta_{s}\right)}\lim_{n\to0}\partial_{n}\int d\mu_{s}\left(w_{s}\right)\prod_{a=1}^{n}d\mu_{t}\left(w_{t}^{a}\right)e^{-\beta_{s}\mathcal{L}_{s}\left(w_{s}\right)}\nonumber \\
 & e^{-\beta_{t}\sum_{a=1}^{n}\mathcal{L}_{t}\left(\left\{ w_{t}^{a}\right\} \right)-\frac{\gamma}{2}\sum_{a}\left\Vert w_{s}-w_{t}^{a}\right\Vert ^{2}}\,.\label{eq:SI-FP_replica}
\end{align}
Eq. \eqref{eq:SI-FP_replica} represents a single-instance generalization
of the classic disordered-averaged Franz-Parisi approach, originally
developed to study metastable states in spin-glasses. We can employ
$f$ to compute the posterior average of relevant quantities, such
as training error, weight norms and distance across weight matrices,
using simple differentiation:

\begin{align}
 & \mathcal{\epsilon}_{t}\equiv\left\langle \mathcal{L}_{t}\right\rangle =-\frac{N_{1}}{P_{t}}\partial_{\beta_{t}}f\,,\\
 & \left\Vert w_{t}\right\Vert ^{2}=-2N_{1}\partial_{\lambda}f\,,\\
 & \left\Vert w_{s}-w_{t}\right\Vert ^{2}=2N_{1}\partial_{\gamma}f\,.
\end{align}
The calculation of the generalization error is slightly more involved:
we introduce and additional term $\beta_{\tau}\mathcal{L}_{\tau}\left(\left\{ w_{t}\right\} \right)$
in the target Hamiltonian, where $\mathcal{L}_{\tau}\left(\left\{ w_{t}\right\} \right)=\frac{1}{2}\sum_{\mu=1}^{P_{\tau}}\left[y_{\tau}^{\mu}-\phi\left(\theta_{t},x_{\tau}^{\mu}\right)\right]^{2}$
is the loss computed on a test set composed of $P_{\tau}$ patterns,
and evaluate the test error with the expression $\mathcal{\epsilon}_{\tau}\equiv\left\langle \mathcal{L}_{\tau}\right\rangle =-\frac{N_{1}}{P_{\tau}}\partial_{\beta_{\tau}}f|_{\beta_{\tau}=0}$.

\paragraph*{Generalization to a one-hidden layer convolutional neural network}

The output of a one-hidden layer convolutional neural network (CNN)
with filters of size $M$ and stride $S$ can be written as:
\begin{equation}
\phi^{CNN}\left(x;\left\{ w,v\right\} \right)=\frac{1}{\sqrt{N_{c}}}\sum_{i=1}^{N_{0}/S}\sum_{c}^{N_{c}}v_{i}^{c}\sigma\left(\frac{1}{\sqrt{M}}\sum_{m}w_{cm}x_{Si+m}\right)\,.
\end{equation}
Very few changes are in order to generalize our calculation to the
case of a shallow CNN, which we will thus highlight along the way.

\subsection{Notation}

We use the $0$ index for all weights and parameters involving the
source, thus denoting the prior inverse variances in each layer $l$
as $\lambda_{l}^{0}=\lambda_{s,l}$ and $\lambda_{l}^{a}\equiv\lambda_{t,l}$
for $a>0$. We thus treat $\left(a,\mu\right)$ as a multi-index over
a construction with concatenated source and replicated target inputs
$[X_{s},\underbrace{X_{t},...X_{t}}_{n}]$, and trust the context to
make the summation over $\mu$ clear. The same is done for the inverse
temperatures, i.e. we have $\beta_{\mu}^{0}=\beta_{s}$ and $\beta_{\mu}^{a}=\beta_{t}$
for $a>0$. All sums over $a$ will implicitly run from $0$ to $n$,
unless specified by the subscript.

We will denote by $\mathbb{1}_{m}$ the identity matrix in dimension
$m$, the vector $\left(e_{m}^{j}\right)_{i}=\delta_{ij}$ is the
canonical base vector and $\left(\mathbb{I}_{m}\right)_{i}=1$ the
constant vector of value $1$. We will usually drop the subscript
when the dimension $m$ is implied by the context. We denote by $\mathcal{N}\left(h;m,C\right)$
a Gaussian distribution with mean $m$ and covariance $C$ over the
vector $h$. All subleading factors will be discarded to reduce clutter.

\section{Transfer learning in a one-hidden layer networks}

Our aim is to compute the following replicated partition function:
\begin{align}
Z^{n}= & \int\prod_{ki}dw_{ki}e^{-\frac{1}{2}\sum_{a}\lambda_{1}^{a}\left\Vert w^{a}\right\Vert ^{2}-\frac{\gamma}{2}\sum_{a}\left\Vert w^{0}-w^{a}\right\Vert ^{2}}\nonumber \\
 & \int\prod_{ak}dv_{k}^{a}e^{-\frac{1}{2}\sum\lambda_{2}^{a}\left\Vert v^{a}\right\Vert ^{2}-\frac{1}{2}\sum_{a\mu}\beta_{\mu}^{a}\left(\sum_{k}\frac{v_{k}^{a}}{\sqrt{N_{1}}}\sigma\left(\frac{1}{\sqrt{N_{0}}}\sum_{i}w_{ki}^{a}x_{\mu i}^{a}\right)-y_{\mu}^{a}\right)^{2}}\,.
\end{align}
We expect that the final form will read 

\begin{equation}
Z^{n}\sim\int\mathcal{DQ}\mathcal{D\bar{Q}}e^{\frac{N_{1}}{2}nS\left(\mathcal{Q},\bar{\mathcal{Q}}\right)}\,,
\end{equation}
with $\mathcal{Q}$ and $\mathcal{\bar{Q}}$ a set of order parameters
whose values will be determined by saddle-point equations. The $\mathcal{O}\left(1\right)$
component $e^{\frac{N_{1}}{2}S_{s}\left(Q_{s},\bar{Q}_{s}\right)}$
of the replicated partition function $Z^{n}$ only depends
on source order parameters and will cancel out with the term $Z_{s}^{-1}$
at the saddle point.

\subsection{Integrating first layer weights}

Introducing the definition for the first-layer replicated pre-activations
$h_{\mu k}^{a}=\frac{1}{\sqrt{N_{0}}}\sum_{i}w_{ki}^{a}x_{i}^{\mu a}$,
we have:

\begin{align}
Z^{n}= & \int\prod_{ki}w_{ki}e^{-\frac{1}{2}\sum_{a}\lambda_{1}^{a}\left\Vert w\right\Vert ^{2}-\frac{\gamma}{2}\sum_{a}\left\Vert w^{0}-w^{a}\right\Vert ^{2}}\nonumber \\
 & \int\prod_{\mu ak}d\bar{h}_{\mu k}^{a}dh_{\mu k}^{a}e^{i\sum_{a\mu k}\bar{h}_{\mu k}^{a}\left(\bar{h}_{k}^{a}-\frac{1}{\sqrt{N_{0}}}\sum_{i}w_{ki}^{a}x_{\mu i}^{a}\right)}\nonumber \\
 & \int\prod_{ak}dv_{k}^{a}e^{-\frac{\lambda_{2}^{a}}{2}\sum_{a}\left\Vert v^{a}\right\Vert ^{2}-\frac{1}{2}\sum_{a\mu}\beta_{\mu}^{a}\left(\sum_{k}\frac{v_{k}^{a}}{\sqrt{N_{1}}}\sigma\left(h_{\mu k}^{a}\right)-y_{\mu}^{a}\right)^{2}}\,.
\end{align}
Let us isolate the dependence over the first layer pre-activations
$h$ in the function $\psi$:
\begin{align}
\psi\left(h\right) & =\int\prod_{a\mu k}d\bar{h}_{\mu k}^{a}e^{i\sum_{a\mu k}\bar{h}_{\mu k}^{a}h_{\mu k}^{a}}\nonumber \\
 & \int\mathcal{D}w\prod_{ki}e^{-\frac{1}{2}\sum_{ab}\Lambda_{1}^{ab}w_{ki}^{a}w_{ki}^{b}-\frac{i}{\sqrt{N_{0}}}\sum_{a\mu}w_{ki}^{a}\bar{h}_{\mu k}^{a}x_{\mu i}^{a}}
\end{align}
where we have defined the following coupling matrix:
\begin{equation}
\Lambda_{1}=\left(\begin{array}{ccccc}
\lambda_{s,1} & -\gamma & -\gamma & ... & -\gamma\\
-\gamma & \lambda_{t,1} & 0 & ... & 0\\
-\gamma & 0 & \lambda_{t,1} & ... & ...\\
... & ... & ... & ... & 0\\
-\gamma & 0 & ... & 0 & \lambda_{t,1}
\end{array}\right)
\end{equation}
and $\mathcal{D}w\equiv\prod_{aki}dw_{ki}^{a}$. We thus write compactly:
\begin{equation}
Z^{n}=\int\mathcal{D}h\psi\left(h\right)\int\prod_{ak}dv_{l}^{a}e^{-\frac{1}{2}\sum_{a\mu}\beta_{\mu}^{a}\left(\sum_{k}\frac{v_{k}^{a}}{\sqrt{N_{1}}}\sigma\left(h_{\mu k}^{a}\right)-y_{\mu}^{a}\right)^{2}-\frac{\lambda_{2}^{a}}{2}\sum_{a}\left\Vert v^{a}\right\Vert ^{2}}
\end{equation}
and integrate over the weights
\begin{equation}
\psi\left(h\right)=\Delta_{1}\prod_{ki}\int\prod_{a\mu}d\bar{h}_{\mu k}^{a}e^{i\sum_{a\mu k}\bar{h}_{\mu k}^{a}\bar{h}_{\mu k}^{a}-\frac{1}{2}\sum_{ab}\Lambda_{ab}^{-1}\bar{q}_{ki}^{a}\left(\bar{h}\right)\bar{q}_{ki}^{b}\left(\bar{h}\right)}
\end{equation}
with the definitions
\begin{align}
 & \bar{q}_{ki}^{a}\left(\bar{h}\right)=\frac{1}{\sqrt{N_{0}}}\sum_{\mu}\bar{h}_{\mu k}^{a}x_{\mu i}^{a}\\
 & \Delta_{1}=e^{-\frac{N_{1}N_{0}}{2}\log\det\Lambda}\,.
\end{align}
We can write $Z^{n}$ in terms of the zero-mean, Gaussian distributed
variables $h_{\mu k}^{a}$, whose covariance matrices $\tilde{C}$
read, for each $k$:
\begin{equation}
\tilde{C}_{\mu\nu}^{ab}=\left\langle h_{\mu k}^{a}h_{\mu k}^{b}\right\rangle =\Lambda_{ab}^{-1}C_{\mu\nu}^{ab}
\end{equation}
where $C_{\mu\nu}^{ab}$ are replicated input covariances:
\begin{equation}
C_{\mu\nu}^{ab}=\frac{1}{N_{0}}\sum_{i=1}^{N_{0}}x_{\mu i}^{a}x_{\mu i}^{b}\,.
\end{equation}
We thus have:
\begin{equation}
Z^{n}=\Delta_{1}\int\prod_{k}\mathcal{D}h_{k}\mathcal{D}v_{k}\mathcal{N}\left(h_{k};0,\tilde{C}\right)e^{-\frac{\lambda_{2}^{a}}{2}\sum_{a}\left\Vert v^{a}\right\Vert ^{2}-\frac{1}{2}\sum_{a\mu}\beta_{\mu}^{a}\left(\sum_{k}\frac{v_{k}^{a}}{\sqrt{N_{1}}}\sigma\left(h_{\mu k}^{a}\right)-y_{\mu}^{a}\right)^{2}}
\end{equation}
with $\mathcal{D}h_{k}\mathcal{D}v_{k}\equiv\prod_{a\mu}dh_{\mu k}^{a}\prod_{a}dv_{k}^{a}$.

\paragraph*{Generalization to a 1-hl convolutional neural network}

In the case of a convolutional layer, we would operate in the same
manner and find $\bar{q}_{cm}^{a}=\frac{1}{\sqrt{M}}\sum_{\mu i}\bar{h}_{\mu ci}^{a}x_{Si+m}^{\mu}$.
Note that variables $\bar{q}$ carry an additional index $c$ for
each patch, so that $\mathcal{\bar{Q}}_{ab}$ is an $N_{c}\times N_{c}$
dimensional matrices, with $N_{c}$ the number of patches. The inter-patch
input covariances reads:
\begin{align}
 & \tilde{C}_{\mu\nu}^{ab,ij}=\left\langle h_{\mu i}^{a}h_{\nu j}^{b}\right\rangle =\left(\Lambda^{-1}\right)^{ab}C_{\mu\nu}^{ab,ij}\\
 & C_{\mu\nu}^{ab,ij}=\frac{1}{M}\sum_{m}x_{Si+m}^{a,\mu}x_{Sj+m}^{b,\nu}\,.
\end{align}

\subsection{Integrating readout weights}

Introducing the definition of the readout outputs $s_{\mu}^{a}=\sum_{k}\frac{v_{k}^{a}}{\sqrt{N_{1}}}\sigma\left(h_{\mu k}^{a}\right)$
with appropriate $\delta$ functions, we obtain an expression of the
form
\begin{equation}
Z^{n}=\Delta_{1}\int\mathcal{D}s\psi\left(s\right)
\end{equation}
where
\begin{align}
\psi\left(s\right)= & \int\mathcal{D}s\mathcal{D}\bar{s}\prod_{k}\mathcal{D}h_{k}\mathcal{D}v_{k}\mathcal{N}\left(h_{k};0,\tilde{C}\right)\nonumber \\
 & \prod_{k}e^{-\frac{1}{2}\sum_{a}\lambda_{2}^{a}\left(v_{k}^{a}\right)^{2}}e^{-\frac{i}{\sqrt{N_{1}}}\sum_{a\mu}\bar{s}_{\mu}^{a}\sum_{k}v_{k}^{a}\sigma\left(h_{\mu k}^{a}\right)}\nonumber \\
 & e^{i\sum_{a\mu}\bar{s}_{\mu}^{a}s_{\mu}^{a}-\frac{1}{2}\sum_{a\mu}\beta_{\mu}^{a}\left(s_{\mu}^{a}-y_{\mu}^{a}\right)^{2}}
\end{align}
with the shorthand $\mathcal{D}s\mathcal{D}\bar{s}=\prod_{a\mu k}ds_{\mu k}^{a}d\bar{s}_{\mu k}^{a}$.
After a straghtforward integration over the uncoupled second-layer
weights, we have:
\begin{align}
\psi\left(s\right)= & \int\mathcal{D}s\mathcal{D}\bar{s}\prod_{k}\mathcal{D}h_{k}\mathcal{D}v_{k}\mathcal{N}\left(h_{k};0,\tilde{C}\right)\nonumber \\
 & \Delta_{2}\prod_{k}e^{-\frac{i}{N_{1}}\sum_{a\mu}\frac{\bar{s}_{\mu}^{a}\bar{s}_{\nu}^{a}}{\lambda_{2}^{a}}\sigma\left(h_{\mu k}^{a}\right)\sigma\left(h_{\mu k}^{a}\right)}\nonumber \\
 & e^{i\sum_{a\mu}\bar{s}_{\mu}^{a}s_{\mu}^{a}-\frac{1}{2}\sum_{a\mu}\beta_{\mu}^{a}\left(s_{\mu}^{a}-y_{\mu}^{a}\right)^{2}}
\end{align}
where we have introduced $\Delta_{2}=e^{-\frac{N_{1}}{2}\log\Lambda_{2}}$ using the second-layer coupling matrix
\begin{equation}
\Lambda_{2}=\left(\begin{array}{ccccc}
\lambda_{s,2} & 0 & 0 & ... & 0\\
0 & \lambda_{t,2} & 0 & ... & 0\\
0 & 0 & \lambda_{t,2} & ... & ...\\
... & ... & ... & ... & 0\\
0 & 0 & ... & 0 & \lambda_{t,2}
\end{array}\right)\,.
\end{equation}
We are now ready to employ the factorization over the first hidden
layer index $k$ and consider the Gaussian variables $h_{\mu}^{a}$.
Following the same strategy of Refs. \cite{pacelli2023statistical,aiudi2023}, we perform a self-consistent Gaussian approximation on the set of variables $\bar{q}^{a}=\frac{1}{\sqrt{\lambda_{2}^{a}N_{1}}}\sum_{\mu}\bar{s}_{\mu}^{a}\sigma\left(h_{\mu}^{a}\right)$
in replica space, with order-parameter covariance matrix:
\begin{equation}
\mathcal{\bar{Q}}^{ab}=\left\langle \bar{q}^{a}\bar{q}^{b}\right\rangle =\frac{1}{N_{1}\sqrt{\lambda_{2}^{a}\lambda_{2}^{b}}}\sum_{\mu\nu}\bar{s}_{\mu\nu}^{a}K_{\mu\nu}^{ab}\bar{s}_{\mu\nu}^{b}
\end{equation}
and kernels:
\begin{equation}
K_{\mu\nu}^{ab}=\left\langle \sigma\left(h_{\mu}^{a}\right)\sigma\left(h_{\nu}^{a}\right)\right\rangle _{\mathcal{N}\left(h;0,\tilde{C}\right)}\,.
\end{equation}
The relevant replicated kernels are:
\begin{align}
\tilde{K}_{s,\mu\nu}= & \left\langle \sigma\left(h_{\mu}^{0}\right)\sigma\left(h_{\nu}^{0}\right)\right\rangle _{\mathcal{N}\left(h;0,\tilde{C}\right)}\\
K_{st,\mu\nu}= & \left\langle \sigma\left(h_{\mu}^{0}\right)\sigma\left(h_{\nu}^{a}\right)\right\rangle _{\mathcal{N}\left(h;0,\tilde{C}\right)}\qquad a>0\\
K_{t,\mu\nu}= & \left\langle \sigma\left(h_{\mu}^{a}\right)\sigma\left(h_{\nu}^{a}\right)\right\rangle _{\mathcal{N}\left(h;0,\tilde{C}\right)}\qquad a>0\\
K_{tt,\mu\nu}= & \left\langle \sigma\left(h_{\mu}^{a}\right)\sigma\left(h_{\nu}^{b}\right)\right\rangle _{\mathcal{N}\left(h;0,\tilde{C}\right)}\qquad a\neq b;a,b>0
\end{align}
with order parameters:
\begin{align}
 & \bar{Q}_{s}=\bar{\mathcal{Q}}^{00}\\
 & \bar{Q}_{st}=\mathcal{\bar{Q}}^{0a}\quad a>0\\
 & \bar{Q}_{t}=\mathcal{\bar{Q}}^{aa}\quad\;a>0\\
 & \bar{Q}_{tt}=\mathcal{\bar{Q}}^{ab}\quad a\neq b;\;a,b>0\,.
\end{align}
Note that, at this stage, the kernels explicitly depend on the replica
number $n$. Introducing the definitions of the order parameters with
the help of appropriate $\delta$ functions and conjugate parameters
$\mathcal{Q}$, we finally obtain:
\begin{align}
Z^{n} & =\Delta_{1}\Delta_{2}\int\mathcal{D}\mathcal{Q}\mathcal{D}\bar{\mathcal{Q}}e^{\frac{N_{1}}{2}\Tr\left(\tilde{\Lambda}_{2}\mathcal{Q}\mathcal{\bar{Q}}\right)-\frac{1}{2}\log\det\left(\mathbb{1}+\mathcal{\bar{Q}}\right)}\nonumber \\
 & \int\mathcal{D}s\mathcal{D}\bar{s}e^{i\bar{s}^{T}s-\frac{1}{2}\bar{s}^{T}\mathcal{K}\bar{s}-\frac{1}{2}\left(s-y\right)^{T}B\left(s-y\right)}
\end{align}
where $\mathcal{K}$ is the renormalized kernel $\mathcal{K}_{\mu\nu}^{ab}=\mathcal{Q}^{ab}K_{\mu\nu}^{ab}$
and $\left(\tilde{\Lambda}_{2}\right)_{ab}=\sqrt{\lambda_{2}^{a}\lambda_{2}^{b}}$.
To simplify notation, we introduced a replicated vector $\left(s\right)_{a\mu}=s_{\mu}^{a}$
for the readout variables and targets $y_{\mu}^{a}$, and correspondingly for the diagonal matrix
$B_{\mu\nu}^{ab}$ containing the inverse temperatures. More concretely,
we have:
\begin{equation}
B=\left(\begin{array}{ccccc}
\beta_{s}\mathbb{1}_{P_{s}} & 0 & 0 & ... & 0\\
0 & \beta_{t}\mathbb{1}_{P_{t}} & 0 & ... & 0\\
0 & 0 & \beta_{t}\mathbb{1}_{P_{t}} &  & ...\\
... & ... & ... & ... & ...\\
0 & 0 & 0 &  & \beta_{t}\mathbb{1}_{P_{t}}
\end{array}\right)
\end{equation}
Integrating over $\bar{s}$ is straightforward:
\begin{align}
Z^{n} & =\Delta_{1}\Delta_{2}e^{-\frac{1}{2}y^{T}By}\int\mathcal{D}\mathcal{Q}\mathcal{D}\mathcal{\bar{Q}}e^{\frac{N_{1}}{2}\Tr\left(\tilde{\Lambda}_{2}\mathcal{Q}\mathcal{\bar{Q}}\right)-\frac{1}{2}\log\det\left(\mathbb{1}+\mathcal{\bar{Q}}\right)}\nonumber \\
 & e^{-\frac{1}{2}s^{T}\left(B+\mathcal{K}^{-1}\right)s+s^{T}By-\frac{1}{2}\log\det\mathcal{K}}\,.
\end{align}
Further integrating over $s$, we find the final form valid for integer number replicas
\begin{equation}
Z^{n}=\int\mathcal{D}\mathcal{Q}\mathcal{D}\bar{\mathcal{Q}}e^{\frac{N_{1}}{2}S\left(\mathcal{Q},\mathcal{\bar{Q}}\right)}
\end{equation}
where the finite-$n$ action is given by:

\begin{align}
S & =-N_{0}\log\det\Lambda_{1}-\log\det\Lambda_{2}+\Tr\left(\tilde{\Lambda}_{2}\mathcal{Q}\mathcal{\bar{Q}}\right)-\log\det\left(\mathbb{1}+\mathcal{\bar{Q}}\right)\nonumber \\
 & -\frac{1}{N_{1}}\log\det{B \Sigma}-\frac{1}{N_{1}} y^{T} \Sigma^{-1}y,
\end{align}
with
\begin{equation}
\Sigma=B^{-1}+\mathcal{K}
\end{equation}

\paragraph*{Generalization to a one-hidden layer convolutional neural network}

In the case of a convolutional network, we would operate in the
same manner and find $\bar{q}_{0,cm}^{CNN,a}=\frac{1}{\sqrt{M}}\sum_{\mu i}\bar{h}_{\mu ci}^{a}x_{Si+m}^{\mu}$,
from which the replicated renormalized local kernel $\mathcal{K}_{\mu\nu}^{ab}=\sum_{ij}\mathcal{Q}_{ij}^{ab}K_{ij,\mu\nu}^{ab}$ is found.

\subsection{Source and target action}

As explained at the beginning of this section, the source-only action
$S_{s}$ can be easily obtained from the $\mathcal{O}\left(1\right)$
terms in $n$:

\begin{align}
S_{s} & =\lambda_{s,2}Q_{s}\bar{Q}_{s}-\log\left(1+\bar{Q}_{s}\right)-\frac{1}{N_{1}}\log\det\beta_{s}\Sigma_{s}-\frac{1}{N_{1}}y_{s}{}^{T}\Sigma_{s}^{-1}y_{s}+\nonumber \\
 & -N_{0}\log\det\lambda_{s,1}-\log\det\lambda_{s,2}
\end{align}
with 
\begin{equation}
\Sigma_{s}=\frac{\mathbb{1}}{\beta_{s}}+Q_{s}K_{s}
\end{equation}
and $K_{s}$ the source-only first-layer kernel:
\begin{equation}
K_{s,\mu\nu}=\left\langle \sigma\left(h_{\mu}\right)\sigma\left(h_{\nu}\right)\right\rangle _{\mathcal{N}\left(h;0,\frac{C_{s}}{\lambda_{s,1}}\right)}
\end{equation}
The values of $Q_{s}$, $\bar{Q}_{s}$ are taken from the source-only
saddle-point equations involved in determining $Z_{s}$, as in the
classic Franz-Parisi approach.

The genuine transfer action is obtained collecting the $\mathcal{O}\left(n\right)$
terms. The algebraic details involved in the diagonalization of the
$\mathcal{Q}$ and $\Sigma$ matrices are collected in section \ref{par:Useful-algebraic-relations}.
Introducing the modified kernel matrices
\begin{align}
 & \Sigma_{t}=\frac{\mathcal{\mathbb{1}}}{\beta_{t}}+Q_{t}K_{t}\\
 & \Delta\Sigma_{t}=\Sigma_{t}-Q_{tt}K_{tt}
\end{align}
and taking the $n\to0$ limit, we finally have:
\begin{equation}
S=\Psi_{T}-\Psi_{\Delta}-\frac{1}{N_{1}}\Psi_{S}-\frac{1}{N_{1}}\Psi_{\mathcal{L}}
\end{equation}
where
\begin{align}
 & \Psi_{T}=2\sqrt{\lambda_{s,2}\lambda_{t,2}}\bar{Q}_{st}Q_{st}+\lambda_{t,2}\left(\bar{Q}_{t}Q_{t}-\bar{Q}_{tt}Q_{tt}\right)\nonumber \\
 & -\log\left(1+\bar{Q}_{t}-\bar{Q}_{tt}\right)-\frac{\left(1+\bar{Q}_{s}\right)\bar{Q}_{tt}-\bar{Q}_{st}^{2}}{\left(1+\bar{Q}_{s}\right)\left(1+\bar{Q}_{t}-\bar{Q}_{tt}\right)}\\
 & \Psi_{\Delta}=N_{0}\left(\log\tilde{\lambda}+\frac{\lambda_{t,1}\gamma}{\tilde{\lambda}\lambda_{s,1}}\right)+\log\det\lambda_{t,2}\\
 & \Psi_{S}=\log\det\beta_{t}\Delta\Sigma_{t}+Q_{tt}\Tr\left(\Delta\Sigma_{t}^{-1}K_{tt}\right)-Q_{st}^{2}\Tr\left(\Sigma_{s}^{-1}K_{st}\Delta\Sigma_{t}^{-1}K_{st}^{T}\right)\\
 & \Psi_{\mathcal{L}}=y_{t}^{T}\Delta\Sigma_{t}^{-1}y_{t}^{T}-2Q_{st}y_{s}^{T}\Sigma_{s}^{-1}K_{st}\Delta\Sigma_{t}^{-1}y_{t}+Q_{st}^{2}y_{s}^{T}\Sigma_{s}^{-1}K_{st}\Delta\Sigma_{t}^{-1}K_{st}^{T}\Sigma_{s}^{-1}y_{s}
\end{align}
with $\tilde{\lambda}=\lambda_{t,1}+\gamma$.
The three relevant kernels for the target action are the following:
\begin{align}
 & K_{st,\mu\nu}=\left\langle \sigma\left(h_{\mu}\right)\sigma\left(h_{\nu}\right)\right\rangle _{\mathcal{N}\left(h;0,\tilde{C}_{st}\right)}\label{eq:K_st_1hl}\\
 & K_{t,\mu\nu}=\left\langle \sigma\left(h_{\mu}\right)\sigma\left(h_{\nu}\right)\right\rangle _{\mathcal{N}\left(h;0,\tilde{C}_{t}\right)}\label{eq:K_t_1hl}\\
 & K_{tt,\mu\nu}=\left\langle \sigma\left(h_{\mu}\right)\sigma\left(h_{\nu}\right)\right\rangle _{\mathcal{N}\left(h;0,\tilde{C}_{tt}\right)}\label{eq:K_tt_1hl}
\end{align}
where the source-target and target modified covariance matrices are
given by
\begin{align}
 & \tilde{C}_{st}=\frac{\gamma}{\tilde{\lambda}\lambda_{s,1}}C_{st}\label{eq:Ctilde_st_1hl}\\
 & \tilde{C}_{t}=\frac{1}{\tilde{\lambda}}\left(1+\frac{\gamma^{2}}{\tilde{\lambda}\lambda_{s,1}}\right)C_{t}\label{eq:Ctilde_t_1hl}
\end{align}
and the inter-replica target kernel is computed using a matrix $\tilde{C}_{tt}$
with entries 
\begin{equation}
\tilde{C}_{tt,\mu\nu}=\delta_{\mu\nu}\tilde{C}_{t,\mu\mu}+\left(1-\delta_{\mu\nu}\right)\frac{\gamma^{2}}{\tilde{\lambda}^{2}\lambda_{s,1}}C_{t,\mu\nu}\label{eq:Ctilde_tt_1hl}
\end{equation}
where $\tilde{\lambda}\equiv\lambda_{t,1}+\gamma$. In the
case of erf activation, we use the well known expression for the NNGP kernel \cite{computing96}
\begin{equation}
\left\langle \sigma\left(h_{\mu}\right)\sigma\left(h_{\nu}\right)\right\rangle _{\mathcal{N}\left(h;0,C\right)}=\frac{2}{\pi}\arcsin\left(\frac{2C_{\mu\nu}}{\sqrt{\left(1+2C_{\mu\mu}\right)\left(1+2C_{\nu\nu}\right)}}\right).
\end{equation}

\subsection{SP equations}

The SP equations for the order parameters read
\begin{align}
 & \sqrt{\lambda_{s,2}\lambda_{t,2}}Q_{st}+\frac{\bar{Q}_{st}}{\left(1+\bar{Q}_{s}\right)\left(1+\bar{Q}_{t}-\bar{Q}_{tt}\right)}=0\\
 & \lambda_{t,2}Q_{t}=\frac{\bar{Q}_{st}^{2}+\left(1+\bar{Q}_{s}\right)\left(1+\bar{Q}_{t}-2\bar{Q}_{tt}\right)}{\left(1+\bar{Q}_{s}\right)\left(1+\bar{Q}_{t}-\bar{Q}_{tt}\right)^{2}}\\
 & \lambda_{t,2}Q_{tt}=\frac{\bar{Q}_{st}^{2}-\left(1+\bar{Q}_{s}\right)\bar{Q}_{tt}}{\left(1+\bar{Q}_{s}\right)\left(1+\bar{Q}_{t}-\bar{Q}_{tt}\right)^{2}}
\end{align}
whereas for the conjugate variables we have:
\begin{align}
 & N_{1}\sqrt{\lambda_{s,2}\lambda_{t,2}}\bar{Q}_{st}=-Q_{st}\Tr\left(\Sigma_{s}^{-1}K_{st}\Delta\Sigma_{t}^{-1}K_{st}^{T}\right)+Q_{st}y_{s}^{T}\Sigma_{s}^{-1}K_{st}\Delta\Sigma_{t}^{-1}K_{st}^{T}\Sigma_{s}^{-1}y_{s}\nonumber \\
 & -Q_{st}y_{s}^{T}\Sigma_{s}^{-1}K_{st}\Delta\Sigma_{t}^{-1}y_{t}\\
\nonumber \\
 & N_{1}\lambda_{t,2}\bar{Q}_{t}=\Tr\left(\Delta\Sigma_{t}^{-1}K_{t}\right)-Q_{tt}\Tr\left(\Delta\Sigma_{t}^{-1}K_{t}\Delta\Sigma_{t}^{-1}K_{tt}\right)\nonumber \\
 & +Q_{st}^{2}\Tr\left(\Sigma_{s}^{-1}K_{st}\Delta\Sigma_{t}^{-1}K_{t}\Delta\Sigma_{t}^{-1}K_{st}^{T}\right)\nonumber \\
 & -Q_{st}^{2}y_{s}^{T}\Sigma_{s}^{-1}K_{st}\Delta\Sigma_{t}^{-1}K_{t}\Delta\Sigma_{t}^{-1}K_{st}^{T}\Sigma_{s}^{-1}y_{s}+2Q_{st}y_{s}^{T}\Sigma_{s}^{-1}K_{st}\Delta\Sigma_{t}^{-1}K_{t}\Delta\Sigma_{t}^{-1}y_{t}\nonumber \\
 & -y_{t}^{T}\Delta\Sigma_{t}^{-1}K_{t}\Delta\Sigma_{t}^{-1}y_{t}^{T}\\
\nonumber \\
 & N_{1}\lambda_{t,2}\bar{Q}_{tt}=-Q_{tt}\Tr\left(\Delta\Sigma_{t}^{-1}K_{tt}\Delta\Sigma_{t}^{-1}K_{tt}\right)+Q_{st}^{2}\Tr\left(\Sigma_{s}^{-1}K_{st}\Delta\Sigma_{t}^{-1}K_{tt}\Delta\Sigma_{t}^{-1}K_{st}^{T}\right)+\nonumber \\
 & -Q_{st}^{2}y_{s}^{T}\Sigma_{s}^{-1}K_{st}\Delta\Sigma_{t}^{-1}K_{tt}\Delta\Sigma_{t}^{-1}K_{st}^{T}\Sigma_{s}^{-1}y_{s}+2Q_{st}y_{s}^{T}\Sigma_{s}^{-1}K_{st}\Delta\Sigma_{t}^{-1}K_{tt}\Delta\Sigma_{t}^{-1}y_{t}\nonumber \\
 & -y_{t}^{T}\Delta\Sigma_{t}^{-1}K_{tt}\Delta\Sigma_{t}^{-1}y_{t}^{T}\,.
\end{align}

\subsection{Norm of the weights and generalization error}

The norm of the second-layer weights can be easily obtained differentiating
the action with respect to $\lambda_{t,2}$:
\begin{equation}
N_{1}\left\langle \left\Vert v\right\Vert ^{2}\right\rangle =\frac{1}{\lambda_{t,2}}+Q_{tt}\bar{Q}_{tt}-Q_{t}\bar{Q}_{t}-\sqrt{\frac{\lambda_{s,2}}{\lambda_{t,2}}}Q_{st}\bar{Q}_{st}
\end{equation}
As explained in section \ref{subsec:Setting-and-notation}, we can
compute the generalization error by differentiating the free-energy
with respect to an additional fictitious temperature $\beta_{\tau}$,
coupled with a loss computed on a test set. This can be easily done
by carrying out the previous calculation with an extended target dataset
obtained by concatenating the input matrices $X_{\tilde{t}}=\left[X_{t},X_{\tau}\right]$
and output vectors $y_{\tilde{t}}=\left[y_{t},y_{\tau}\right]$. Accordingly,
each target block in the matrix $B$ is extended so as to contain
$P_{t}+P_{\tau}$ diagonal entries from the vector $\beta_{\mu}=(\underbrace{\beta_{t},...,\beta_{t}}_{P_{t}},\underbrace{\beta_{\tau},...,\beta_{\tau}}_{P_{\tau}})$.
In what follows, the kernels $K_{s\tilde{t}}$, $K_{\tilde{t}}$ and
$K_{\tilde{t}\tilde{t}}$ are obtained from equations (\ref{eq:K_st_1hl}--\ref{eq:Ctilde_tt_1hl})
using covariance matrices from the extended dataset $X_{\tilde{t}}$.
Calling $\Delta K=Q_{t}K_{\tilde{t}}-Q_{tt}K_{\tilde{t}\tilde{t}}$
and further denoting its test-test and train-test blocks respectively
as $\Delta K_{\tau}$ and $\Delta K_{t\tau}$, we finally have for
the generalization error:
\begin{align}
\epsilon_{\tau} & =\frac{1}{2}\Tr\left(\Delta K_{\tau}-\Delta K_{t\tau}^{T}\Delta\Sigma_{t}^{-1}\Delta K_{t\tau}\right)+\frac{1}{2}\left\Vert y_{\tau}-\Delta K_{t\tau}^{T}\Delta\Sigma_{t}^{-1}y_{t}\right\Vert ^{2}+\nonumber \\
 & \frac{1}{2}Q_{tt}\Tr\left(\tilde{M}K_{\tilde{t}\tilde{t}}\right)-\frac{1}{2}Q_{st}^{2}\Tr\left(\Sigma_{s}^{-1}K_{s\tilde{t}}\tilde{M}K_{s\tilde{t}}^{T}\right)+\nonumber \\
 & \frac{1}{2}Q_{st}^{2}y_{s}^{T}\Sigma_{s}^{-1}K_{s\tilde{t}}\tilde{M}K_{s\tilde{t}}^{T}\Sigma_{s}^{-1}y_{s}-Q_{st}y_{s}^{T}\Sigma_{s}^{-1}K_{s\tilde{t}}\tilde{M}y_{\tilde{t}}
\end{align}
where the matrix $\tilde{M}$ reads:
\begin{equation}
\tilde{M}=\left(\begin{array}{cc}
\Delta\Sigma_{t}^{-1}\Delta K_{t\tau}\Delta K_{t\tau}^{T}\Delta\Sigma_{t}^{-1} & -\Delta\Sigma_{t}^{-1}\Delta K_{t\tau}\\
-\Delta K_{t\tau}^{T}\Delta\Sigma_{t}^{-1} & \mathbb{1}_{P_{\tau}}
\end{array}\right).
\end{equation}

\subsection{Some algebraic details in our derivation}

\paragraph{Useful algebraic relations.\label{par:Useful-algebraic-relations}} We gather here some algebraic relations involving a block-matrix $\mathcal{F}$,
useful in both layer-wise integration of weights coupling and the
ensuing calculations. Given four matrices $A,B,C,\Delta C$ with $A\in \mathbb{R}^{p_{1}\times p_{1}}$, $B\in \mathbb{R}^{p_{1}\times p_{2}}$ and $C, \Delta C \in \mathbb{R}^{p_{2}\times p_{2}}$, we call $\mathcal{F}$ the $\left(p_{1}+np_{2}\right)\times\left(p_{1}+np_{2}\right)$
dimensional block matrix of the form:
\begin{equation}
\mathcal{F}=\left(\begin{array}{cccc}
A & B & B & B\\
B^{T} & C+\Delta C & C & C\\
B^{T} & C & C+\Delta C & C\\
B^{T} & C & C & C+\Delta C
\end{array}\right)\,.\label{eq:our_friend_matrix}
\end{equation}
We can easily compute its inverse by first considering its lower-right
$n\times n$ block

\begin{equation}
\mathcal{F}_{n}=\mathbb{1}_{n}\otimes\Delta C+\mathbb{I}_{n}\mathbb{I}_{n}^{T}\otimes C
\end{equation}
where:

\begin{align}
 & \mathcal{F}_{n}^{-1}=\mathbb{1}_{n}\otimes\Delta C^{-1}+\mathbb{I}_{n}\mathbb{I}_{n}^{T}\otimes\bar{\mathcal{F}}\\
 & \bar{\mathcal{F}}=\frac{1}{n}\left[\left(\Delta C+nC\right)^{-1}-\Delta C^{-1}\right]=-\left(\Delta C+nC\right)^{-1}C\Delta C^{-1}\,.
\end{align}
We have for $\mathcal{F}^{-1}$ the block structure:
\begin{equation}
\mathcal{F}^{-1}=\left(\begin{array}{cc}
\mathcal{F}_{0}^{-1} & \mathbb{I}_{n}\otimes\mathcal{F}_{1}\\
\mathbb{I}_{n}^{T}\otimes\mathcal{F}_{1}^{T} & \mathcal{F}_{2,n}^{-1}
\end{array}\right)
\end{equation}
with the Schur complements and off-diagonal term reading:
\begin{align}
 & \mathcal{F}_{0}=A-nB\left(\Delta C+nC\right)^{-1}B^{T}\\
 & \mathcal{F}_{1}=-A^{-1}B\left(\Delta C+nC-nB^{T}A^{-1}B\right)^{-1}\\
 & \mathcal{F}_{2,n}=\mathbb{1}_{n}\otimes\Delta C+\mathbb{I}_{n}\mathbb{I}_{n}^{T}\otimes\left(C-B^{T}A^{-1}B\right)\\
 & \mathcal{F}_{2,n}^{-1}=\mathbb{1}_{n}\otimes\Delta C^{-1}+\mathbb{I}_{n}\mathbb{I}_{n}^{T}\otimes\frac{1}{n}\left[ \left(\Delta C+n\bar{\mathcal{F}}_{2}\right)^{-1}-\Delta C^{-1}\right] \\
 & \bar{\mathcal{F}}_{2}=C-B^{T}A^{-1}B\,.
\end{align}
As for the determinant of $\mathcal{F}$ one has:

\begin{align}
 & \log\det\mathcal{F}_{n}=\left(n-1\right)\log\det\Delta C+\Tr\log\left(\Delta C+nC\right)\\
 & \log\det\mathcal{F}=\log\det\mathcal{F}_{n}+\Tr\log\left(\mathcal{F}^{0}\right)\,.
\end{align}
We will consider the contraction of $\mathcal{F}$ with a vector of
the form $\tilde{y}=(y_{0},\underbrace{y,...,y}_{n})$:
\begin{equation}
\tilde{y}^{T}\mathcal{F}^{-1}\tilde{y}=y_{0}^{T}\mathcal{F}_{0}^{-1}y_{0}+2ny_{0}^{T}\mathcal{F}_{1}y+ny^{T}\left(\Delta C+n\bar{\mathcal{F}}_{2}\right)^{-1}y
\end{equation}
Deriving the final action for transfer learning involves the $n\to0$
limits of the previously computed quantities:
\begin{align}
 & \log\det\mathcal{F}_{n}\sim n\log\det\Delta C+n\Tr\left(\Delta C^{-1}C\right)\\
 & \Tr\log\mathcal{F}^{0}\sim \Tr\log A-n\Tr\left(A^{-1}B\Delta C^{-1}B^{T}\right)\\
 & \mathcal{F}_{0}^{-1}\sim A^{-1}+nA^{-1}B\Delta C^{-1}B^{T}A^{-1}\\
 & \mathcal{F}_{1}\sim-A^{-1}B\left[\Delta C^{-1}-n\Delta C^{-1}\left(C-B^{T}A{}^{-1}B\right)\Delta C^{-1}\right]\\
 & y^{T}\mathcal{F}^{-1}y\sim y_{0}^{T}\mathcal{F}_{0}^{-1}y_{0}+2ny_{0}^{T}\mathcal{F}_{1}y+ny^{T}\Delta C^{-1}y
\end{align}

\paragraph{Layer-wise integration over weights}

The coupling over the replicated weights involves an un-normalized
Gaussian with zero and inverse covariance $\Lambda$
\begin{equation}
\varphi\left(w;\Lambda\right)=\prod_{ij}e^{-\sum_{a}\Lambda_{ab}w_{ij}^{a}w_{ij}^{b}}
\end{equation}
such that integration over the replicated weights $w_{ij}^{a}$ takes
the form:
\begin{equation}
\int\mathcal{D}w\varphi\left(w;\Lambda\right)\prod_{ki}e^{-iw^{T}\bar{q}}=\prod_{ki}\det\Lambda e^{-\frac{1}{2}\bar{q}_{ki}^{T}\Lambda^{-1}\bar{q}_{ki}}.
\end{equation}
In particular, we model transfer learning with an $\left(n+1\right)\times\left(n+1\right)$
matrix:
\begin{equation}
\Lambda=\left(\begin{array}{cc}
\lambda_{s}+\gamma n & -\gamma\mathbb{I}_{n}^{T}\\
-\gamma\mathbb{I}_{n} & \tilde{\lambda}\mathbb{1}_{n}
\end{array}\right)
\end{equation}
with the notation $\tilde{\lambda}=\lambda_{t}+\gamma$. Calling $c=\lambda_{t}\lambda_{s}-n\gamma^{2}$,
its inverse and determinant read:
\begin{align}
 & \Lambda^{-1}=\left(\begin{array}{cc}
\frac{\tilde{\lambda}}{c} & \frac{\gamma}{c}\mathbb{I}_{n}^{T}\\
\frac{\gamma}{c}\mathbb{I}_{n} & D^{-1}
\end{array}\right)\\
 & \log\det\Lambda=\left(n-1\right)\log\tilde{\lambda}+\log c\sim n\log\left(\lambda_{t}+\gamma\right)+n\frac{\lambda_{t}\gamma}{\left(\lambda_{t}+\gamma\right)\lambda_{s}}+\log\lambda_{s}
\end{align}
with:
\begin{align}
 & D=\tilde{\lambda}\mathbb{1}_{n}-\frac{\gamma^{2}}{\lambda_{s}+\gamma n}\mathbb{I}_{n}\mathbb{I}_{n}^{T}\\
 & D^{-1}=\frac{1}{\tilde{\lambda}}\left(\mathbb{1}_{n}+\frac{\gamma^{2}}{\tilde{\lambda}\lambda_{s}+n\lambda_{t}\gamma}\mathbb{I}_{n}\mathbb{I}_{n}^{T}\right)
\end{align}

\paragraph{Details on determinants and quadratic forms}

The determinant of the $\left(n+1\right)\times\left(n+1\right)$ matrix
$\det\left(\mathbb{1}_{n+1}+\bar{\mathcal{Q}}\right)$ can be easily
worked out by using formulas from the previous section: 
\begin{align}
 & \det\left(\mathbb{1}_{n+1}+\mathcal{\bar{Q}}\right)=\det\bar{\mathcal{Q}}_{n}\left(1+\bar{Q}_{s}-\bar{Q}_{st}^{2}\mathbb{I}_{n}^{T}\bar{\mathcal{Q}}_{n}^{-1}\mathbb{I}_{n}\right)=\nonumber \\
 & \left(1+\bar{Q}_{t}-\bar{Q}_{tt}\right)^{n-1}\left(1+\bar{Q}_{t}-\bar{Q}_{tt}+n\bar{Q}_{tt}\right)\left(1+\bar{Q}_{s}-\frac{n\bar{Q}_{st}^{2}}{1+\bar{Q}_{t}-\bar{Q}_{tt}+n\bar{Q}_{tt}}\right)
\end{align}
In the small $n$ limit it has the form:
\begin{align}
 & \log\det\left(\mathbb{1}_{n+1}+\mathcal{\bar{Q}}\right)=\log\left(1+\bar{Q}_{s}\right)+\nonumber \\
 & n\left[\log\left(1+\bar{Q}_{t}-\bar{Q}_{tt}\right)+\frac{\left(1+\bar{Q}_{s}\right)\bar{Q}_{tt}-\bar{Q}_{st}^{2}}{\left(1+\bar{Q}_{s}\right)\left(1+\bar{Q}_{t}-\bar{Q}_{tt}\right)}\right]\,.
\end{align}
We recall that the replicated coupling matrix for last-layer pre-activations reads:
\begin{equation}
\Sigma=\left(\begin{array}{ccccc}
\tilde{\Sigma}_{s} & Q_{st}K_{st} & Q_{st}K_{st} & ... & Q_{st}K_{st}\\
Q_{st}K_{st}^{T} & \Sigma_{t} & Q_{tt}K_{tt} & ... & Q_{tt}K_{tt}\\
Q_{st}K_{st}^{T} & Q_{tt}K_{tt} & \Sigma_{t} & ... & ...\\
... & ... & ... & ... & Q_{tt}K_{tt}\\
Q_{st}K_{st}^{T} & Q_{tt}K_{tt} & ... & Q_{tt}K_{tt} & \Sigma_{t}
\end{array}\right)
\end{equation}
with $\tilde{\Sigma}_{s}=\frac{\mathbb{1}}{\beta_{s}}+Q_{s}K_{s}$. In the expansion of $\log\det{\Sigma}$ and the quadratic form $y^{T}\Sigma^{-1}y$,
we encounter the terms
\begin{equation}
\tilde{\Sigma}_{s}=\left(\frac{\mathbb{1}}{\beta_{s}}+Q_{s}\left(K_{s}+n\delta K_{s}\right)\right)^{-1}\sim\Sigma_{s}^{-1}-nQ_{s}\Sigma_{s}^{-2}\delta K_{s}
\end{equation}
arising from $\tilde{K}_{s,\mu\nu}=\left\langle \sigma\left(h_{\mu}\right)\sigma\left(h_{\nu}\right)\right\rangle \sim K_{s,\mu\nu}+n\delta K_{s,\mu\nu}$.
We will however drop the $\mathcal{O}\left(n\right)$ terms since
they do not contribute to the SP equations for the transfer order
parameters. We thus have for the two quantities:
\begin{align}
 & \log\det\left(B^{-1}+\Sigma\right)\sim \Tr\log\Sigma_{s}+n\log\det\Delta\Sigma_{t}+nQ_{tt}\Tr\left(\Delta\Sigma_{t}^{-1}K_{tt}\right)\nonumber \\
 & -nQ_{st}^{2}\Tr\left(\Sigma_{s}^{-1}K_{st}\Delta\Sigma_{t}^{-1}K_{st}^{T}\right)\\
 & y^{T}\left(B^{-1}+\Sigma\right)^{-1}y\sim y_{0}^{T}\Sigma_{s}^{-1}y_{0}+nQ_{st}^{2}y_{0}^{T}\Sigma_{s}^{-1}K_{st}\Delta\Sigma_{t}^{-1}K_{st}^{T}\Sigma_{s}^{-1}y_{0}+\nonumber \\
 & -2nQ_{st}y_{0}^{T}\Sigma_{s}^{-1}K_{st}\Delta\Sigma_{t}^{-1}y+ny^{T}\Delta\Sigma_{t}^{-1}y\,.
\end{align}

\section{Transfer learning in a simple regression problem}

In this short supplementary section, we address the transfer learning
problem in the simplest possible setting, linear regression for both
the \emph{source} and the \emph{target} task.

\subsection{Derivation}

The learning problem for the source vector $w_{s}$ is defined by
the minimization of the regularized training loss:
\begin{equation}
\mathcal{L}_{s}=\frac{1}{2}\sum_{\mu=1}^{P_{s}}\left(y_{s}^{\mu}-\sum_{i}\frac{w_{s,i}x_{s,i}^{\mu}}{\sqrt{N}}\right)^{2}+\frac{N\lambda_{s}}{2}\sum_{i=1}^{N}w_{s,i}^{2}
\end{equation}
which yields the known solution:
\begin{equation}
w_{s}=\left(X_{s}^{T}X_{s}+N\lambda_{s}\mathbb{1}_{N}\right)^{-1}X_{s}^{T}y_{s}\,.
\end{equation}
With the help of the Woodbury identity we get from the previous expression:
\begin{equation}
w_{s}=\frac{1}{N}X_{s}^{T}G_{s}y_{s}
\end{equation}
with the definition
\begin{equation}
G_{s}=\left(\lambda_{s}\mathbb{1}_{P_{s}}+C_{s}\right)^{-1}\Leftrightarrow\mathbb{1}_{P_{s}}-G_{s}C_{s}=\lambda_{s}G_{s}.
\end{equation}
The transfer learning problem for the target weight vector $w_{t}$
is defined by the minimization of the regularized training loss in
the presence of a source-target coupling:
\begin{equation}
\mathcal{L}_{t}=\frac{1}{2}\sum_{\mu=1}^{P_{t}}\left(y_{t}^{\mu}-\sum_{i}\frac{w_{t,i}x_{t,i}^{\mu}}{\sqrt{N}}\right)^{2}+\frac{N\lambda_{t}}{2}\sum_{i=1}^{N}w_{t,i}^{2}+\frac{N\gamma}{2}\sum_{i=1}^{N}\left(w_{t,i}-w_{s,i}\right)^{2}
\end{equation}
yielding:
\begin{equation}
w=\left(X_{t}^{T}X_{t}+N\tilde{\lambda}\mathbb{1}_{N}\right)^{-1}\left(X^{T}y_{t}+N\gamma w_{s}\right)
\end{equation}
with the notation $\tilde{\lambda}=\lambda_{t}+\gamma$. Again, by
calling $G_{t}=\left(\tilde{\lambda}\mathbb{1}_{P_{t}}+C_{t}\right)^{-1}$,
we have:
\begin{equation}
\left(\frac{X_{t}^{T}X_{t}}{N}+\tilde{\lambda}\mathbb{1}_{N}\right)^{-1}=\frac{1}{\tilde{\lambda}}\left(\mathbb{1}_{N}-\frac{1}{N}X_{t}^{T}G_{t}X_{t}\right)\,.
\end{equation}
Introducing the notations
\begin{align}
 & \tilde{y}_{s}=\lambda_{s}G_{s}y_{s}\\
 & \tilde{y}_{t}=\tilde{\lambda}G_{t}y_{t}\\
 & \tilde{Y}=\left[\tilde{y}_{s},\tilde{y}_{t}\right]
\end{align}
we write succintly:
\begin{equation}
w_{t}=\frac{1}{N\tilde{\lambda}}\left(\begin{array}{c}
\frac{\gamma}{\lambda_{s}}\left(X_{s}^{T}-X_{t}^{T}G_{t}C_{st}^{T}\right)\\
X_{t}^{T}
\end{array}\right)\tilde{Y}
\end{equation}
where we have defined $C_{st}=\frac{X_{s}X_{t}^{T}}{N}$.

\paragraph{Norm of target weight vector}

The norm of $w_{t}$ is easily written as:
\begin{equation}
\left\Vert w_{t}\right\Vert ^{2}=\frac{1}{N\tilde{\lambda}^{2}}\tilde{Y}\tilde{\mathcal{W}}\tilde{Y}^{T}
\end{equation}
defining the matrix:
\begin{align}
 & \mathcal{\tilde{W}}=\left(\begin{array}{cc}
\mathcal{\tilde{W}}_{s} & \mathcal{\tilde{W}}_{st}\\
\mathcal{\tilde{W}}_{st}^{T} & \mathcal{\tilde{W}}_{t}
\end{array}\right)\\
 & \mathcal{\tilde{W}}_{s}=\frac{\gamma^{2}}{\lambda_{s}^{2}}\left[C_{s}-C_{st}G_{t}\left(\mathbb{1}_{P_{t}}+\tilde{\lambda}G_{t}\right)C_{st}^{T}\right]\\
 & \mathcal{\tilde{W}}_{st}=\frac{\gamma\tilde{\lambda}}{\lambda_{s}}C_{st}G_{t}\\
 & \tilde{\mathcal{W}}_{t}=C_{t}\,.
\end{align}

\paragraph{Training and test error}

Inserting the solution $w_{t}$ in the expression for the training
error we get, after some manipulation:
\begin{equation}
\epsilon_{t}=\frac{1}{2}\left\Vert y_{t}-X_{t}w_{t}\right\Vert ^{2}=\frac{1}{2}\frac{\gamma^{2}}{\lambda_{s}^{2}}\tilde{y}_{s}^{T}C_{st}G_{t}^{2}C_{st}^{T}\tilde{y}_{s}-\frac{\gamma}{\lambda_{s}}\tilde{y}_{s}^{T}C_{st}G_{t}\tilde{y}_{t}+\frac{1}{2}\tilde{y}_{t}^{T}\tilde{y}_{t}\,.
\end{equation}
The error over the test set is given by the following:
\begin{align}
\epsilon_{g} & =\frac{1}{2}\left\Vert y_{\tau}-X_{\tau}w_{t}\right\Vert ^{2}=\nonumber \\
 & \frac{\gamma^{2}}{2\lambda_{s}^{2}\tilde{\lambda}^{2}}\tilde{y}_{s}^{T}G_{s\tau}G_{s\tau}^{T}\tilde{y}_{s}+\nonumber \\
 & \frac{1}{2\tilde{\lambda}^{2}}\tilde{y}^{T}C_{t\tau}C_{t\tau}^{T}\tilde{y}_{t}+\frac{\gamma}{\lambda_{s}\tilde{\lambda}^{2}}\tilde{y}_{s}^{T}G_{s\tau}C_{t\tau}^{T}\tilde{y}_{t}\nonumber \\
 & -\frac{\gamma}{\lambda_{s}\tilde{\lambda}}y_{\tau}^{T}G_{s\tau}^{T}\tilde{y}_{s}-\frac{1}{\tilde{\lambda}}y_{\tau}^{T}C_{t\tau}^{T}\tilde{y}_{t}+\frac{1}{2}y_{\tau}^{T}y_{\tau}
\end{align}
with the source-test and target-test covariances defined respectively
as $C_{s\tau}=\frac{X_{s}X_{\tau}^{T}}{N}$ and $C_{t\tau}=\frac{X_{t}X_{\tau}^{T}}{N}$
and $G_{s\tau}=C_{s\tau}-C_{st}G_{t}C_{t\tau}$.

\subsection{A simple example of transfer regression}

\begin{figure}
\includegraphics[scale=0.8]{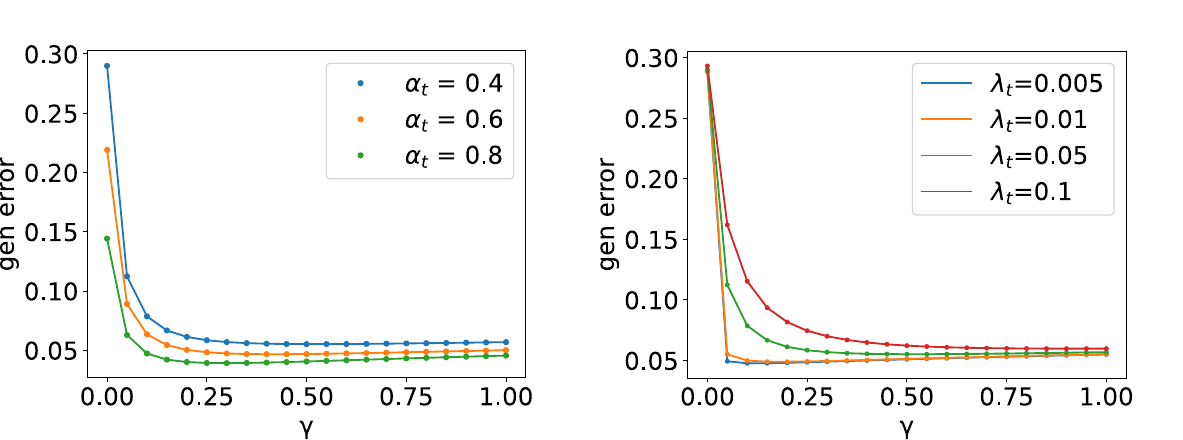}\caption{\label{fig:Transfer-regression}Transfer learning in a linear regression
task defined by a random teacher vector $w_{0}$. In both panels curves
are theoretical results, points are obtained by solving the source
and target problem with standard optimization methods. Parameters:
$N=200$, $P_{\tau}=10.000$ \textbf{A}: Generalization error as a
function of the coupling parameter $\gamma$ for different sizes of
the training set for the target problem, indicated by the ratio $\alpha_{t}=P_{t}/N$.\textbf{
B}: Same as \textbf{A} for different values of the regularization
$\lambda_{t}$ of the target weight vector $w_{t}$.}
\end{figure}

In Figure \ref{fig:Transfer-regression} we show  an example of a transfer
learning problem in a single-layer, linear version of the task shown
in Fig. 2 of the main text, in the special case with two identical
teachers for the source and target tasks. We generate the three datasets
$X_{s}$, $X_{t}$ and $X_{t}$ with normal i.i.d. entries. The task
is defined in terms of a random teacher weight vector $w_{0}$ with
zero-mean and Gaussian i.i.d. entries with standard deviation $1/N$
(we use $N=200$ in this examples): outputs for the three sets are
linear functions of the inputs $X_{\alpha}$ given by $y_{\alpha}=X_{\alpha}w_{0}$
for $\alpha\in\left\{ s,t,\tau\right\} $. The source weight vector
$w_{s}$ is trained using $P_{s}=\alpha_{s}N$ data-points, with $\alpha_{s}=0.8$.
The transfer effect is apparent in the decrease of the generalization
error as a function of the source-target coupling parameter $\gamma$.

\section{Transfer learning in a deep fully connected network}

In this section, we sketch a tentative derivation of the effective action in the case of networks with $L$ hidden layers. In this case, all the layers except for the last are coupled among source and target, while the readout ($L+1$ in our convention) remains uncoupled. The replicated priors for the weights at each layer are defined by the coupling matrices $\Lambda_l$:

\begin{equation}
\Lambda_{l}=\left(\begin{array}{ccccc}
\lambda_{s,l} & -\gamma & -\gamma & ... & -\gamma\\
-\gamma & \lambda_{t,l} & 0 & ... & 0\\
-\gamma & 0 & \lambda_{t,l} & ... & ...\\
... & ... & ... & ... & 0\\
-\gamma & 0 & ... & 0 & \lambda_{t,l}
\end{array}\right)\,,\qquad\Lambda_{L+1}=\left(\begin{array}{ccccc}
\lambda_{s,L+1} & 0 & 0 & ... & 0\\
0 & \lambda_{t,L+1} & 0 & ... & 0\\
0 & 0 & \lambda_{t,L+1} & ... & ...\\
... & ... & ... & ... & 0\\
0 & 0 & ... & 0 & \lambda_{t,L+1}
\end{array}\right)\,.
\end{equation}

We are interested in the computation of the replicated partition function $Z^n$, using index $a = 1 \ldots n$ for replicas. To describe a fully connected layer with weight matrix $w_l$, we introduce the following notation for the pre-activations at layer $l+1$:
\begin{equation}
h_{l+1}^a=w^a_{l}x^a_{l},\qquad x^a_{l}=\sigma\left(h^a_{l}\right) \,.   
\end{equation}
To easily deal with the recursion over layer, let us introduce a definition for the function computed by a neural network between any two intermediate layers $l$, $l'$:
\begin{align}
h_{l'}^a=\phi_{l:l'}^a\left( h_{l};\left\{ w^a\right\} _{l+1:l'}\right)\,.
\end{align}
The function $\phi_{l:l'}$ takes as inputs the
pre-activations at layer $l$ and outputs the pre-activations at
layer $l'$, with the convention that $h_{0}=x$. We have stressed the dependence of $\phi_{l:l'}$ on all the weights of the layers between $l$ and $l'$, $\{w\}_{l+1:l'}$, but we will immediately drop it to ease the notation. The output of the network reads $ \phi\left(x\right)\equiv\phi_{0:L+1}\left(x \right)$, and we can write the loss in terms of the pre-activations of layer $l$ as:
\begin{equation}
\mathcal{L}\left(\left\{ w\right\} \right)=\frac{1}{2}\sum_{\mu}\left(\phi\left(x^{\mu}\right)-y^{\mu}\right)^{2}=\sum_{\mu}\left(\phi_{l:L+1}\left(h_{l}^{\mu}\right)-y^{\mu}\right)^{2}\,.
\end{equation}
Let us define:
\[
\chi_{l:l'}\left(h_{l}^{\mu}; \{w\}_{l+1:l'} \right)=\exp\left[-\frac{1}{2}\sum_{\mu a}\beta_{a}\left(\phi_{l:l'}\left(h_{l}^{\mu}\right)-y_{\mu}^{a}\right)^{2}\right] \, ,
\]
%
Using this convention we can rewrite the free-entropy as:
\begin{align}
f & =\frac{1}{N_{L}}\frac{1}{Z_{s}\left(\beta_{s}\right)}\lim_{n\to0}\partial_{n}Z^{n}  \\
Z^{n} &=\int\prod_{l}\mathcal{D}  w_{l} \, \chi_{0:L+1} (x) e^{-\frac{1}{2}\sum_{l}w_{l}^{T}\Lambda_{l}w_{l}} \, ,
\end{align}
with the shorthands $\mathcal{D}w_l=\prod_{i_{l}i_{l-1},a}dw_{i_{l}i_{l-1}}^{l,a}$ and $w_{l}^{T}\Lambda_{l}w_{l}=\sum_{i_{l}i_{l-1}}\sum_{ab}w_{l,i_{l}i_{l-1}}^{l,a}\Lambda_{l,ab}w_{l,i_{l}i_{l-1}}^{l,b}$.

\subsection{Integrating first-layer weights}

The integration over the first-layer weights $w^a_{1,i_0i_1}$ is straightforward, since
for each $i_{1},i_{0}$ the coupling is Gaussian with inverse covariance
$\Lambda_{1}$. Let us isolate the integral over the first-layer weights and consider the dependence over the first layer pre-activations:
\begin{equation}
Z^n=\int\prod_{l>1}\mathcal{D}w_{l}e^{-\frac{1}{2}\sum_{l>1}w_{l}^{T}\Lambda_{l}w_{l}}\int\mathcal{D}h_1\chi_{1:L+1}\left(h_{1}\right)\psi\left(h_1\right)
\end{equation}
where:
\begin{equation}
\psi\left(h_{1}\right)=\prod_{i_{1}}\int\prod_{\mu}d\bar{h}_{1,\mu i_{1}}^{a}e^{i\sum_{a\mu}\bar{h}_{1,\mu i_{1}}^{a}h_{1,\mu i_{1}}^{a}}\prod_{i_{0}}\int dwe^{-\frac{1}{2}\sum_{ab}w^{a}\Lambda_{1}^{ab}w^{b}-\frac{i}{\sqrt{N_{0}}}\sum_{a\mu}w^{a}\bar{h}_{1,\mu i_{1}}^{a}x_{\mu i_{0}}^{a}}.
\end{equation}
Integrating over the weights and using the notation $
\Delta_{l}=\mathrm{exp}\{-\frac{1}{2}N_{l-1}N_{l}\log\det\Lambda_{l}\}$, we have:
\begin{equation}
\psi\left(h_{1}\right)=\Delta_{1}\prod_{i_{1}i_{0}}\int\prod_{\mu}d\bar{h}_{1,\mu i_{1}}^{a}e^{i\sum_{a\mu}\bar{h}_{1,\mu i_{1}}^{a}\bar{h}_{1,\mu i_{1}}^{a}-\frac{1}{2}\sum_{ab}\Lambda_{ab}^{-1}\bar{q}_{0,i_{1}i_{0}}^{a}\left(\bar{h}\right)\bar{q}_{0,i_{1}i_{0}}^{b}\left(\bar{h}\right)}
\end{equation}
with the definition
\begin{equation}
\bar{q}_{0,i_{1}i_{0}}^{a}\left(\bar{h}\right)=\frac{1}{\sqrt{N_{0}}}\sum_{\mu}\bar{h}_{\mu i_{1}}^{a}x_{\mu i_{0}}^{a}\,.
\end{equation}
Employing the factorization over the first layer index $i_{1}$ and summing over $i_{0}$, we can write the replicated partition function as follows
\begin{equation}
Z^{n}=\Delta_{1}\int\prod_{l>1}\mathcal{D}w_{l}e^{-\frac{1}{2}\sum_{l>1}w_{l}^{T}\Lambda_{l}w_{l}}\int\prod_{i_{1}}\left\{ \mathcal{D}h_{1,i_{1}}\mathcal{N}\left(h_{1,i_{1}};0,\Sigma_{1}\right)\right\} \chi_{1:L+1}\left(\left\{ h_{1}\right\} _{i_{1}}\right)
\end{equation}
where, for each $i_1$, the replicated pre-activations are Gaussian with covariance matrices $\Sigma_1$:
\begin{equation}
\tilde{C}_{\mu\nu}^{ab}=\left\langle h_{1,\mu i_{1}}^{a}h_{1,\nu i_{1}}^{b}\right\rangle =\left(\Lambda_{1}^{-1}\right)^{ab}C_{\mu\nu}^{ab}
\end{equation}
in turn depending on the replicated input covariances:
\begin{equation}
C_{\mu\nu}^{ab}=\frac{1}{N_{0}}\sum_{i_{0}=1}^{N_{0}}x_{\mu i_{0}}^{a}x_{\nu i_{0}}^{b}\,.
\end{equation}

\subsection{Recursion relation for deep FC network}

Introducing the definition for the second-layer pre-activations we
have:
\begin{equation}
Z^n=\Delta_{1}\int\prod_{l>2}\mathcal{D}w_{l}e^{-\frac{1}{2}\sum_{l>2}w_{l}^{T}\Lambda_{l}w_{l}}\int\mathcal{D}h_2\chi_{2:L+1}\left(h_2\right)\psi\left(h_2\right)   
\end{equation}
with:
\begin{align}
\psi\left(h_{2}\right) & =\int\mathcal{D}\bar{h}_{2}e^{i\bar{h}_{2}^{T}h_{2}}\int\prod_{i_{1}}\left\{ \mathcal{D}h_{1,i_{1}}\mathcal{N}\left(h_{1,i_{1}};0,\Sigma_{1}\right)\right\} \nonumber \\
 & \int\mathcal{D}w_{2}\prod_{i_{2}i_{1}}e^{-\frac{i}{\sqrt{N_{1}}}\sum_{a}w_{2,i_{2}i_{1}}^{a}\sum_{\mu}\bar{h}_{2,\mu i_{2}}^{a}\sigma\left(h_{1,\mu i_{1}}^{a}\right)}
\end{align}
We again introduce the quantities
\begin{equation}
    \bar{q}_{1,i_{2}i_{1}}^{a}\left(\bar{h}_{2}\right)=\frac{1}{\sqrt{N_{1}}}\sum_{\mu}\bar{h}_{2,\mu i_{2}}^{a}\sigma\left(h_{1,\mu i_{1}}^{a}\right)
\end{equation}
and perform the integration over the weights $w_2$. Employing the factorization over the index $i_1$ (and dropping the index for clarity) we write
\begin{equation}
\psi\left(h_{2}\right)=\Delta_{2}\prod_{i_{2}i_{1}}\int\mathcal{D}\bar{h}_{2}e^{i\bar{h}_{2}^{T}h_{2}}\left(\int\mathcal{D}\bar{q}_{1}\varphi\left(\bar{q}_{1}\right)e^{-\frac{1}{2}\bar{q}_{1}^{T}\bar{q}_{1}-\frac{1}{2}\bar{q}_{1}^{T}\Lambda_{2}^{-1}\bar{q}_{1}}\right)^{N_{1}}
\end{equation}
with:
\begin{equation}
\varphi\left(\bar{q}_{1}\right)=\left\langle \prod_{i_{2}}\delta\left(\bar{q}_{1,i_{2}}^{a}-\frac{1}{\sqrt{N_{1}}}\sum_{\mu}\bar{h}_{2,\mu i_{2}}^{a}\sigma\left(h_{1,\mu}^{a}\right)\right)\right\rangle _{h_{1}}
\end{equation}
To deal with the extensive extensive $N_{2}n$ number of variables
$\bar{q}_{1}$, we employ a self-consistent Gaussian approximation
with covariance matrix
\begin{equation}
\bar{\mathcal{Q}}_{1,i_{2}j_{2}}^{ab}=\left\langle \bar{q}_{1,i_{2}}^{a}\bar{q}_{1,j_{2}}^{b}\right\rangle =\frac{1}{N_{1}}\sum_{\mu\nu}\bar{h}_{2,\mu i_{2}}^{a}K_{1,\mu\nu}^{ab}\bar{h}_{2,\nu j_{2}}^{b}
\end{equation}
and kernels
\begin{equation}
K_{1,\mu\nu}^{ab}=\left\langle \sigma\left(h_{1,\mu}^{a}\right)\sigma\left(h_{1,\nu}^{ab}\right)\right\rangle _{\mathcal{N}\left(h_{1};0,\tilde{C}\right)}
\end{equation}
thus getting:
\begin{align}
\left\langle e^{-\frac{1}{2}\bar{q}_{1}^{T}\Lambda_{2}^{-1}\bar{q}_{1}}\right\rangle _{\varphi\left(\bar{q}_{1}\right)} & =e^{-\frac{N_{1}}{2}\Tr_{i_{2},a}\log\left(\mathbb{1}+\left(\Lambda_{2}^{-1}\right)^{ab}\bar{\mathcal{Q}}_{1,i_{2}j_{2}}^{ab}\right)}\sim\nonumber \\
 & e^{-\frac{N_{1}}{2}\Tr_{a}\log\left(\mathbb{1}+\left(\Lambda_{2}^{-1}\right)^{ab}\Tr_{i_{2}}\bar{\mathcal{Q}}_{1,i_{2}j_{2}}^{ab}\right)}
\end{align}
The second line of the previous equations implement a mean-field,
permutation symmetric approximation, whereby we obtained an inter-replica
covariance by tracing over the $N_{2}$ second-layer hidden units.

Introducing the definitions for the new mean-field inter-replica covariance
with appropriate $\delta$ functions, we thus get: 
\begin{align}
\psi\left(h_{2}\right) & =\Delta_{2}\int\mathcal{D}\mathcal{\bar{Q}}_{1}e^{-\frac{1}{2}\log\det\left(\mathbb{1}+\Lambda_{2}^{-1}\mathcal{\bar{Q}}_{1}\right)}\nonumber \\
 & \int\mathcal{D}\bar{h}e^{i\bar{h}^{T}h_{2}}\delta\left(N_{1}\mathcal{\bar{Q}}_{1}^{ab}-\left(\bar{h}^{a}\right)^{T}K_{1}^{ab}\bar{h}^{b}\right)
\end{align}
Expanding the $\delta$'s and easily integrating over $\bar{h}$ we find:
\begin{equation}
\psi\left(h_2\right)=\Delta_2 \int\mathcal{D}\mathcal{Q}_{1}\mathcal{D}\bar{\mathcal{Q}}_{1}e^{\frac{N_{1}}{2}\Tr\left(\mathcal{Q}_{1}\mathcal{\bar{Q}}_{1}\right)-\frac{1}{2}\log\det\left(\mathbb{1}+\Lambda_{2}^{-1}\mathcal{\bar{Q}}_{1}\right)}\mathcal{N}\left(h_2;0,\mathcal{K}_{2}\right)
\end{equation}
with
\begin{equation}
\mathcal{K}_{2}^{ab}=\mathcal{Q}_{1}^{ab}K_{1}^{ab}
\end{equation}
The matrix $\mathcal{Q}_{1}$ acts as renormalization for the kernel
$K_{1}$, whereby $h_2$ are Gaussian conditioning on $\mathcal{Q}_{1}$.
Using a simple recursion across layer one gets:
\begin{align}
Z^{n} & =\prod_{l=1}^{L}\Delta_{l}\int\prod_{l=1}^{L-1}\mathcal{D}\mathcal{Q}_{l}\mathcal{D}\bar{\mathcal{Q}}_{l}e^{\frac{1}{2}\sum_{l=1}^{L-1}N_{l}\Tr\left(\mathcal{Q}_{l}\mathcal{\bar{Q}}_{l}\right)-\frac{1}{2}\sum_{l=1}^{L-1}\log\det\left(\mathbb{1}+\Lambda_{l+1}^{-1}\mathcal{\bar{Q}}_{l}\right)}\nonumber \\
 & \int\mathcal{D}ve^{-\frac{1}{2}v^{T}\Lambda_{L+1}v}\int\mathcal{D}h^{L}\chi_{L:L+1}\left(h^{L}\right)\psi\left(h^{L}\right)
\end{align}
with the notation $v\equiv w_{L+1}$ on the uncoupled last layer weights.

\subsection{Readout layer}

Introducing the definition $s=\phi_{L:L+1}\left(h_{L};v\right)$ for
the readout outputs, we obtain a form 
\begin{equation}    
Z^{n}=\prod_{l=1}^{L}\Delta_{l}\int\prod_{l=1}^{L-1}\mathcal{D}\mathcal{Q}_{l}\mathcal{D}\bar{\mathcal{Q}}_{l}e^{\frac{1}{2}\sum_{l=1}^{L-1}N_{l}\Tr\left(\mathcal{Q}_{l}\mathcal{\bar{Q}}_{l}\right)-\frac{1}{2}\sum_{l=1}^{L-1}\log\det\left(\mathbb{1}+\Lambda_{l+1}^{-1}\mathcal{\bar{Q}}_{l}\right)}\int\mathcal{D}s\psi\left(s\right)
\end{equation}
with
\begin{align}
\psi\left(s\right) & =\Delta_{L+1}\int\mathcal{D}\bar{s}e^{i\bar{s}^{T}s-\frac{1}{2}\left(s-y\right)^{T}B\left(s-y\right)}\int\mathcal{D}h_{L}\mathcal{N}\left(h_{L};0,\mathcal{K}_{L}\right)\nonumber \\
 & \int\mathcal{D}v\prod_{i_{L}}e^{-\frac{i}{\sqrt{N_{L}}}\sum_{a}v_{i_{L}}^{a}\sum_{\mu}\bar{s}_{\mu}^{a}\sigma\left(h_{L,\mu i_{L}}^{a}\right)}
\end{align}
We again introduce the variables $\bar{q}^{a}=\frac{1}{\sqrt{N_{L}}}\sum_{\mu}\bar{s}_{\mu}^{a}\sigma\left(h_{L,\mu}^{a}\right)$, where we dropped the index $i_L$ owning to factorization. Employing the usual Gaussian equivalence,
their joint distribution is a normal distribution with order-parameter covariance
matrix
\begin{equation}
\bar{\mathcal{Q}}_{L}^{ab}=\left\langle \bar{q}^{a}\bar{q}^{b}\right\rangle =\frac{1}{N_{L}}\left(\bar{s}^{a}\right)^{T}K_{L}^{ab}\bar{s}^{b}
\end{equation}
and kernels
\begin{equation}
K_{L,\mu\nu}^{ab}=\left\langle \sigma\left(h_{\mu}^{a}\right)\sigma\left(h_{\nu}^{a}\right)\right\rangle _{\mathcal{N}\left(h_{L};0,\mathcal{K}_{L}\right)}
\end{equation}
We finally obtain
\begin{align}
\psi\left(s\right) & =\Delta_{L+1}\int\mathcal{D}\mathcal{Q}_{L}\mathcal{D}\bar{\mathcal{Q}}_{L}e^{\frac{N_{L}}{2}\Tr\left(\mathcal{Q}_{L}\mathcal{\bar{Q}}_{L}\right)-\frac{1}{2}\log\det\left(\mathbb{1}+\Lambda_{L+1}^{-1}\mathcal{\bar{Q}}_{L}\right)}\nonumber \\
 & \int\mathcal{D}\bar{s}e^{i\bar{s}^{T}s-\frac{1}{2}\bar{s}^{T}\mathcal{K}_{L+1}\bar{s}-\frac{1}{2}\left(s-y\right)^{T}B\left(s-y\right)}
\end{align}
where $\mathcal{K}_{L+1}^{ab}=\mathcal{Q}_{L}^{ab}K_{L}^{ab}$.
We now employ the same steps as in the case of the 1hl network, thus arriving at the final form
\begin{equation}
Z^n=\int\prod_{l}\mathcal{D}\mathcal{Q}_{l}\mathcal{D}\bar{\mathcal{Q}}_{l}e^{\frac{N_{L}}{2}S\left(\mathcal{Q},\mathcal{\bar{Q}}\right)}
\end{equation}
with the action
\begin{align}
S & =\sum_{l}^{L}\left[\log\Delta_{l}+\frac{N_{l}}{N_{L}}\Tr\left(\mathcal{Q}_{l}\mathcal{\bar{Q}}_{l}\right)-\frac{N_{l}}{N_{L}}\sum_{l}\log\det\left(\mathbb{1}+\Lambda_{l+1}^{-1}\mathcal{\bar{Q}}_{l}\right)\right]+\nonumber \\
 & \log\Delta_{L+1}-\log\det\left(\mathbb{1}+B\mathcal{K}_{L+1}\right)-y^{T}\left(B^{-1}+\mathcal{K}_{L+1}\right)^{-1}y\,.
\end{align}
The $n\to0$ limit can again be carried out using the expressions for
the determinants and quadratic forms in section \ref{par:Useful-algebraic-relations}. It is worth noticing that we expect this derivation to be exact for deep linear networks, as long as replica symmetry is not broken.

\section{Details of numerical experiments}
In this section, we provide some details about training algorithm and tasks used for numerical validation of our theoretical results. All experiments are performed on pairs of source/target architectures with one hidden layer and Erf (error function) activation. To ensure sampling from the posterior Gibbs ensemble of weights, which is essential to validate our theory, we train our networks using a discretised Langevin dynamics, similarly to what is done in \cite{SompolinskyLinear,seroussi2023natcomm, pacelli2023statistical, baglioni2024predictive}. At each training step $t$, the parameters $\theta = \lbrace w, v\rbrace $ are updated according to: 
\begin{equation}
    \theta(t+1) =  \theta(t) - \eta \nabla_\theta \tilde{\mathcal L}(\theta(t)) +\sqrt{2T\eta}\epsilon(t)
\end{equation}
where $T=1/\beta$ is the temperature, $\eta$ is the learning rate, $\epsilon(t)$ is a white Gaussian noise vector with entries drawn from a standard normal distribution, and the regularized loss function $\tilde{\mathcal L}$ comprises the prior rescaled by the temperature:
\begin{equation}
    \tilde{\mathcal L}_{s/t} = \mathcal{L}_{s/t} + \frac{T\lambda_{s/t,1}}{2} \Vert w_{s/t} \Vert^2 + \frac{T\lambda_{s/t,2}}{2} \Vert v_{s/t} \Vert^2.
\end{equation}
Temperature and learning rate are fixed to $T =10^{-2}$ and $ \eta =  10^{-3}$ throughout the experiments for both source and target networks.

\subsection{Experimental Setup}
We use pairs of correlated source/task classification tasks. Two pairs involve real-world computer vision datasets (C-EMNIST and C-CIFAR), and one is a synthetic task (CHMM). We train the source model on the source task first, then extract $k$ equilibrium configurations of the source weights. For each of the $k$ sets of features, we train a target network for different values of the parameter $\gamma$ controlling the coupling to source weights, and average results over the $k$ source configurations. In Fig.~1 of the main text $k = 5$, in Fig. 2 we used $k = 10$. 

\subsubsection{C-EMNIST and C-CIFAR}
Similarly to~\cite{gerace2022probing}, we build a binary source task by dividing a subset of the EMNIST letters into two distinct groups: letters $\{A, B, E, L\}$ for the first one and $\{C, H, J, S\}$ for the second one. We assign the label to each image according to the group membership. The target task is then built from the source task by replacing one letter per group (letter $E$ with $F$ and $J$ with $I$). We call this pair of source/target tasks C-EMNIST (correlated EMNIST). C-CIFAR is constructed in a similarly way from the CIFAR10 dataset: the source task includes the first $8$ classes, specifically $\{1,2,3,4\}$ in the first group and $\{5,6,7,8\}$ in the second one. The target task is obtained by replacing class $1$ with $10$ and class $5$ with $9$). 

In all experiments, images from CIFAR10 and EMNIST are gray-scaled and down-sized. In Fig. 2 and 3 of the main text, the input size is set to $N_0 = 784$ pixels. in Fig. 1B, the curves at $N_1 = 500$ have $N_0 = 784$, while those at $N_1 = 1000$ are obtained by pre-processing the input data points $x$ with random features:

\begin{equation}
    \hat{x} = \sigma \left(\frac{Fx}{N_0} \right)
\end{equation}
where $F \in \mathbb{R}^{D \times N_0}$ is the random feature matrix, whose entries are sampled iid from a standard Gaussian. This effectively projects the input data points in a new space of dimension $D$. In the experiments in fig.~1B we set $D = 400$.

\subsubsection{CHMM}
To analyze the extent to which the correlation between source/target tasks is essential for TL to be beneficial, we use the \emph{correlated hidden manifold} (CHMM), a synthetic data model where source-target correlations can be explicitly tuned via three different set of parameters, meant to mimic different and realistic TL scenarios \cite{gerace2022probing}.

For instance, the source and the target set may differ because of the traits characterizing the input data point. In the model, this is described via the parameters $\rho$ and $\eta$ which control, respectively, how many features of the source data are replaced by new ones in the target set and how much the remaining features differ between the two tasks. The source-target datasets may instead share the same set of input data but these data could be labelled according to different labelling rules. The mismatch between the labelling rules is controlled via the parameter $q$. Finally, the source-target datasets may differ in terms of the dimensionality of the data manifold, which can be controlled by directly tuning the source and target intrinsic dimensions. 

\end{document}